\documentclass{JHEP3}

  \usepackage{latexsym,bm,amsmath,amssymb,amsfonts}
  \usepackage{epsfig,graphics,graphicx}
  \usepackage{slashed}

  \long\def\comment#1{ }

  \newcommand{\mcal}{\mathcal}
  \newcommand{\rme}{{\rm e}}
  \newcommand{\rmd}{{\rm d}}   

  \newcommand{\nn}{\nonumber\\}
   
  \newcommand{\order}[1]{\mcal{O}{(#1)}}
  \newcommand{\calbfA}{{\bm{\mathcal{A}}}}

  \newcommand{\beq}{\begin{eqnarray}}
  \newcommand{\eeq}{\end{eqnarray}}
  
 \def\simge{\mathrel{%
   \rlap{\raise 0.511ex \hbox{$>$}}{\lower 0.511ex \hbox{$\sim$}}}}
\def\simle{\mathrel{
   \rlap{\raise 0.511ex \hbox{$<$}}{\lower 0.511ex \hbox{$\sim$}}}}

\vspace{1.5cm}

\title{\rm \LARGE Deep inelastic scattering off a ${\mathcal N}=4$
SYM plasma at strong coupling}

\author{Y.~Hatta and E. Iancu\\Service de Physique Th\'eorique, CEA Saclay,
  F-91191 Gif-sur-Yvette, France\\
        E-mail: \email
{Yoshitaka.Hatta@cea.fr}, \email{Edmond.Iancu@cea.fr}}
\author{A. H.~Mueller\\Department of Physics, Columbia University, New York, NY
10027, USA\\
        E-mail: \email{amh@phys.columbia.edu}}

\abstract{By using the AdS/CFT correspondence we study the deep inelastic
scattering of an ${\mathcal R}$--current off a ${\mathcal N}=4$
supersymmetric Yang--Mills (SYM) plasma at finite temperature and strong
coupling. Within the supergravity approximation valid when the number of
colors is large, we compute the structure functions by solving Maxwell
equations in the space--time geometry of the $AdS_5$ black hole.
We find a rather sharp transition between  a low energy regime where the
scattering is weak and quasi--elastic, and a high--energy regime where
the current is completely absorbed. The critical energy for this
transition determines the plasma saturation momentum in terms of its
temperature $T$ and the Bjorken $x$ variable: $Q_s=T/x$. These results
suggest a partonic picture for the plasma where all the partons have
transverse momenta below the saturation momentum and occupation numbers
of order one.}

\begin{document}

\section{Introduction}

Over the recent years, there has been increasing evidence, coming from
the experimental results at RHIC and their theoretical interpretations
\cite{Shuryak:2003xe,Gyulassy:2004zy,Heinz:2005zg,Muller:2007rs}, and
also from theoretical studies of the QCD thermodynamics
\cite{Karsch:2006sf,Endrodi:2007tq,Blaizot:2003tw}, that the hadronic
matter produced after a high--energy heavy ion collision may interact
rather strongly, in spite of being in the deconfined phase of QCD and
having a relatively high partonic density. For instance, the success of
theoretical approaches based on hydrodynamics
\cite{Heinz:2005zg,Teaney:2003kp}, which assumes local thermal
equilibrium and vanishing, or small, viscosity, in describing collective
phenomena like elliptic flow \cite{Ackermann:2000tr,Adcox:2002ms},
suggests rapid thermalization and a low viscosity-to-entropy ratio for
the matter produced at RHIC, which are hallmarks of a nearly--ideal
fluid, with strong interactions. Also, the experimental results for the
`jet--quenching parameter' at RHIC \cite{Adler:2003qi,Adams:2003kv},
which is a measure of the rate at which highly energetic partons loose
energy in the surrounding medium, have been interpreted
\cite{Eskola:2004cr,Dainese:2004te} to yield values which are too large
to be explained by weak coupling calculations
\cite{Baier:1996kr,Baier:2000mf} (but this interpretation is not
universally accepted; see, for instance, \cite{Baier:2006fr}).
Furthermore, lattice studies of the QCD thermodynamics give evidence for
a strong coupling behaviour (like the persistence of meson--like bound
states \cite{Umeda:2002vr,Asakawa:2003re,Datta:2003ww,Aarts:2007pk} and
strong deviations from the pressure of an ideal gas of quarks and gluons
\cite{Karsch:2006sf,Endrodi:2007tq}) up to temperatures a few times the
critical temperature for deconfinement. Such conclusions are corroborated
by analytic calculations for the quark--gluon plasma showing that the
weak--coupling expansion is too poorly convergent to be useful in
practice for all temperatures of interest
\cite{Blaizot:2000fc,Andersen:2002ey,Kajantie:2002wa,CaronHuot:2007gq}.

These and similar observations have urged the need for non--perturbative
studies of the hadronic matter at relativistically high temperatures and
densities. While lattice gauge theory is a privileged tool to
non--perturbatively address static properties like the thermodynamics or
the screening masses, its extension towards dynamical problems, like
transport phenomena, dispersion relations, or the high--energy
scattering, remains prohibitively complicated, and new methods are
therefore required to systematically address such problems at strong
coupling. The AdS/CFT correspondence \cite{ADSCFT}, although so far
limited, in its most convincing formulation, to gauge theories which are
`simpler' (in the sense of having more symmetries) than QCD, is the most
promising candidate in that sense.

This method can most easily deal with the large--$N$ limit, with $N$ the
number of colors, where the gauge coupling $g$ is small but the `t Hooft
coupling $\lambda\!=\!g^2N$ is large, in which case the ${\mathcal
N}\!=\!4$ supersymmetric Yang--Mills (SYM) theory can be mapped onto a
weakly--coupled string theory, that can be studied via semi--classical
techniques. Leaving aside the structural differences between the
${\mathcal N}\!=\!4$ SYM theory, which is conformal, and real QCD
--- these differences can be argued to be less important in well--chosen
physical regimes, and, besides, some of them can be incorporated into
extensions of the ${\mathcal N}\!=\!4$ SYM theory (for which, however,
the AdS/CFT correspondence is less firmly established) ---, it is still
not clear whether the aforementioned parametric conditions can be made
consistent with the situation in QCD, where $N\!=\!3$ and
$g\!\sim\!\order{1}$ (giving $\lambda\!\simeq\! 3\div 6$) in the
interesting physical regimes. But even if a detailed, quantitative,
comparison to real QCD (in particular, to the experimental data) would be
premature, it is nevertheless clear that the AdS/CFT approach can provide
valuable information about the non--perturbative behaviour of gauge
theories, which should allow us to better constraint the physical reality
of QCD from the strong--coupling end.

Given these promising features, and the experimental imperatives at RHIC
or LHC, it is not surprising that, over the last few years, there was a
profusion of applications of the AdS/CFT techniques to problems of
interest for high--density QCD. Following early applications to
thermodynamics \cite{Gubser:1996de,Witten:1998zw} and the pioneering
calculation, by Policastro, Son, and Starinets, of the shear viscosity
\cite{Policastro:2001yc,Policastro:2002se}, there was an intense activity
towards computing the jet--quenching parameter
\cite{Liu:2006ug,Armesto:2006zv,Lin:2006au,Liu:2006he}, the energy--loss
of a heavy quark
\cite{Gubser:2006bz,Herzog:2006gh,Herzog:2006se,Caceres:2006dj} or of a
quark--antiquark pair
\cite{Peeters:2006iu,Liu:2006nn,Chernicoff:2006hi,Caceres:2006ta,Argyres:2006vs,Avramis:2006em},
the diffusion rate for a heavy quark
\cite{CasalderreySolana:2006rq,CasalderreySolana:2007qw}, the energy
disturbances due to moving quarks \cite{Gubser:2007xz,Chesler:2007an},
the Debye screening mass \cite{Bak:2007fk,Amado:2007pv}, the production
rate for photons and dileptons \cite{CaronHuot:2006te}, or the Bjorken
expansion and the approach towards thermalization
\cite{Janik:2005zt,Nakamura:2006ih,Kovchegov:2007pq,Kajantie:2007bn} ---
all of that in the context of the strongly--coupled ${\mathcal N}=4$ SYM
plasma at finite temperature (sometimes extended to include a chemical
potential).

Several of the studies mentioned above have been concerned with the
long--range ($\Delta x\gg 1/T$) or large--time ($\Delta t\gg 1/T$)
behaviour of the strongly--coupled plasma, as relevant e.g. for
hydrodynamics, thermalization, or transport phenomena. On the other hand,
in order to study the propagation of `hard' (i.e., highly energetic and
relatively small) probes through the plasma, like jets or electromagnetic
probes, it is essential to have a good understanding of the plasma
structure on {\em short} space--time separations $\ll 1/T$,  much alike
the parton picture in perturbative QCD. Of course, at strong coupling
there is {\em a priori} not clear whether the notion of a `parton' --- in
the sense of a point--like constituent which behaves as quasi--free
during the interaction with the external probe --- makes sense in the
first place, neither if such a `parton', in case it exists, should belong
to an individual `quasiparticle' (a thermal excitations with energies and
momenta of order $T$), or rather is a property of the plasma as a whole.
In other terms, is the partonic distribution of the plasma (again,
assuming that this exists) the direct sum of the respective distributions
for the constituent quasiparticles, with appropriate thermal weights, or
rather is this qualitatively different ?

Such questions are extremely difficult and below we shall not attempt to
answer them in full generality. In particular, it is not yet understood
whether a strongly--coupled gauge plasma admits a quasiparticle picture
on the thermal scale $1/T$, so like the Landau theory of a Fermi liquid,
or the quasiparticle structure of the quark--gluon plasma emerging from
resummations of perturbation theory \cite{Blaizot:2001nr,Blaizot:2003tw}.
Fortunately, however, there is no need to properly understand the
structure of the plasma on this scale $1/T$ so long as we are merely
interested on the corresponding structure on much shorter space--time
scales $\ll 1/T$. Indeed, the latter can be directly measured (at least,
in a Gedankenexperiment) by an external probe with high energy and
momentum ($\omega,\,q\gg T$). From the experience with QCD we know that
the most convenient measurement of that type --- that whose results are
most directly related to the parton structure of the target --- is the
`deep inelastic scattering' (DIS) of a leptonic probe off the plasma.

DIS at strong coupling in the context of the AdS/CFT correspondence has
been so far considered \cite{dis,Hatta:2007he} only for the case where
the target is a single hadron (a `dilaton'). In this approach, the
`electromagnetic' probe which initiates the scattering is the conserved
current associated with a particular $U(1)$ symmetry (the `${\mathcal
R}$--current'), whose associated `${\mathcal R}$--charge' is also carried
by the light degrees of freedom which are present inside the hadrons. By
computing the current--current correlator in the hadron wavefunction, one
can extract the same information about hadronic structure functions that
would be obtained by DIS via a `photon' coupled to the $U(1)$ current. In
terms of the standard kinematical variables $Q^2$ and $x$, with
$Q^2\!=q^2\!-\omega^2$ the virtuality of the current and $x\approx Q^2/s$
(at high energy $s\gg Q^2$), the DIS structure function $F_2(x,Q^2)$ is a
measure of the number of partons which carry a longitudinal momentum
fraction $x$ and occupy an area $\sim 1/Q^2$ in the transverse, impact
parameter, space.

In this paper, we shall use the same general setup --- the scattering
between the ${\mathcal R}$--current and the plasma --- to compute the
structure functions of a strongly--coupled ${\mathcal N}\!=\!4$ SYM
plasma at finite temperature. It turns that, in this case, the formalism
is quite different --- in fact, somewhat simpler and also conceptually
clearer --- than in the case of a single--hadron target considered in
Ref. \cite{dis,Hatta:2007he}. There are several reasons for such
differences:

First, the string theory dual of the ${\mathcal N}\!=\!4$ SYM plasma is
unambiguously known, at the level of the original formulation of the
AdS/CFT correspondence \cite{ADSCFT} : this is a `black--hole' (more
precisely, a non--extremal black three--brane; see Sect. 2 below for
details) in a curved space--time geometry which is asymptotically
$AdS_5\times S^5$. By contrast, in order to accommodate a hadronic state,
the ${\mathcal N}\!=\!4$ SYM theory (which has no confinement) must be
`deformed' in the infrared, in such a way to break down conformal
symmetry. This deformation is not unique and, besides, its dual analog in
the string theory is generally ambiguous.

Second, the interplay between the large--$N$ limit and the high--energy
limit turns out to be much more subtle for a single--hadron target than
for a plasma. This is in turn related to an essential feature of the
strong--coupling problem, which is the deep connection between the
distribution of partons and the issue of unitarity in DIS at high energy.
As explained in Ref. \cite{Hatta:2007he}, at strong coupling, most of the
partons are concentrated in the kinematical region where the scattering
is strong and the unitarity corrections are important (the analog of the
`saturation', or `color glass condensate', region of perturbative QCD
\cite{SATreviews}). This is an important point that we shall try to
motivate here via general arguments, and for which the subsequent
calculations in this paper will provide an explicit realization:

One can heuristically understand this point by extrapolating the picture
of parton evolution in perturbation theory: Partons at large $x$ tend to
radiate and thus drop down at smaller values of $x$. At weak coupling,
the emitted partons are predominantly soft (i.e., they carry only a tiny
fraction $x'\ll 1$ of the longitudinal momentum of their parent partons),
so, even for very high energies, there is still a substantial fraction of
the parton distribution at relatively large values of $x$. These
large--$x$ partons carry almost all of the hadron energy and momentum,
but they are unimportant for high--energy scattering, which is rather
controlled by the bulk of the distribution at small--$x$. At strong
coupling, on the other hand, there is no penalty for the hard emissions;
the distribution of the energy among the child partons after a branching
is essentially democratic, and hence the overall distribution can very
fast degrade, via successive branchings, down to very small values of
$x$. One therefore expects the structure functions at strong coupling to
be concentrated at small values of $x$, but the question is, how small ?
These functions are, of course, constrained by energy--momentum
conservation --- the small--$x$ partons must carry the overall energy and
momentum of the hadron ---, but this constraint (a `sum--rule' on $F_2$)
is not sufficient to determine the parton distribution. A more severe
constraint comes from unitarity: in the kinematical region where the
scattering is strong, in the sense that the scattering amplitude has
reached the unitarity bound, the structure functions are, by definition,
large, and hence the partons exist. Thus, at strong coupling, the search
for the parton distribution is tantamount to understanding the unitarity
problem for DIS.

This is where the large--$N$ limit becomes important: the elementary
scattering amplitude is suppressed\footnote{There is no similar
suppression for the DIS structure functions because the strength $J^2$ of
the ${\mathcal R}$--current increases like $N^2$, due to the color
degrees of freedom of the fields which make up the current.} by a factor
$1/N^2$, so for a {\em single--hadron target} and in the strict
large--$N$ limit ($N\to\infty$ at fixed energy), the scattering can never
become strong, and thus the bulk of the partons cannot be seen. This is
the situation considered in Ref. \cite{dis}, and indeed it has been found
there that the dilaton has no point--like constituents except at
extremely small values of $x$ (for a given resolution $Q^2$), within a
kinematical domain which squeezes exponentially to zero when increasing
$\lambda$.


But the partonic structure of the dilaton reveals itself after relaxing
the large--$N$ limit, as we did in Ref. \cite{Hatta:2007he}. Namely, we
found that, for sufficiently large $Q^2$, the partons are all located in
the strong--scattering region at $x\lesssim x_s(Q^2)$, where
$x_s(Q^2)\simeq \Lambda^2/(N^2Q^2)$ is the `saturation line' (a line in
the kinematical plane $(x,Q^2)$ along which the elementary amplitude is
constant and of order one) and $\Lambda$ is the infrared cutoff which
fixes the size of the dilaton. Moreover, the phase--space distribution of
these partons turns out to be remarkably simple (and somehow reminiscent
of the gluon distribution in the `color glass condensate' at weak
coupling \cite{SATreviews}): there is essentially a single parton of a
given color per unit cell in phase--space. This {\em a posteriori}
legitimates the use of the `electromagnetic' current as a probe of the
parton distribution: in spite of the coupling being strong, the current
can interact only with one parton at a time, and thus it can faithfully
measure the parton number. Since, moreover, partons with very small $x\ll
x_s$ carry only little energy and momentum, it is clear that the hadron
total energy and momentum is concentrated in the partons near the
saturation line $x=x_s(Q^2)$.

Returning, after this long digression, to the plasma problem of current
interest, we note that in this context one can simplify the problem by
using the large--$N$ approximation without loosing the salient features:
for a plasma target, the scattering can be strong even in the large--$N$
limit, because the plasma involves $N^2$ degrees of freedom per unit
volume --- as manifest from the fact that its entropy density scales like
$N^2T^3$ \cite{Gubser:1996de,Witten:1998zw} ---, which compensates for
the $1/N^2$ suppression of the elementary scattering amplitude. Note
that, at this point, and at several other places in the paper, we use a
heuristic language in which the plasma thermal degrees of freedom are
treated as `quasiparticles' with typical energies and momenta of order
$T$, and the overall scattering process is viewed as the sum of
elementary scatterings between these quasiparticles and the ${\mathcal
R}$-current. This language, inspired by the situation at weak coupling,
is admittedly ambiguous at strong coupling, and is used here only to gain
more intuition into mathematical manipulations which by themselves are
free of any ambiguity.

Specifically, in the large--$N$ limit of interest, the scattering between
the ${\mathcal R}$-current and the plasma can be described in the
supergravity approximation, as the propagation of the gravitational
perturbation induced by the current in the background metric of the black
three--brane. The current--current correlator relevant to DIS is then
computed from the action evaluated on the solution to the classical wave
equation --- an imaginary part in this solution being synonymous of
inelasticity in the scattering of the current. The wave dynamics is
non--trivial in only one dimension --- the radial dimension of $AdS_5$,
which plays the role of an `impact parameter' between the current and the
black hole. The relevant wave equation can be formally rewritten as a
Schr\"odinger equation in one spatial dimension (actually, two such
equations, for the longitudinal and transverse waves, respectively).
Then, the dynamics is controlled by the potential in this equation and,
more precisely, by the competition between two important terms: a
`repulsive' term proportional to $Q^2$ which by itself would keep the
wave at the boundary of $AdS_5$ (far away from the horizon of the black
hole), and an `attractive' term, proportional to the energy times the
temperature, which tends to pull the wave towards the black hole. We thus
distinguish between two physical regimes:

\texttt{(i)} At relatively low energy and/or low temperature, such that
$x \gg T/Q$, the repulsive term dominates, and the wave remains confined
near the boundary. (For DIS off the plasma, $x\sim Q^2/qT$, and we recall
that $T/Q\ll 1$ for the physical problem of interest.) In this regime the
scattering is weak and quasi--elastic (the imaginary part in the
classical solution is extremely small, since generated via tunneling
through the potential). Correspondingly, the DIS structure functions are
exponentially small, e.g., $F_2\sim \exp\{-c(xQ/T)^{1/2}\}$, a result
that we interpret as the absence of point--like constituents in the SYM
plasma having $x \gg T/Q$.

\texttt{(ii)} At high enough energy, such that $x\lesssim x_s(Q)\simeq
T/Q$, the attractive term in the potential takes over, and then the wave
escapes inside the bulk of $AdS_5$, until it gets absorbed by the black
hole. This absorption generates a large imaginary part in the solution,
and hence a large contribution to the structure functions for DIS, which
for $x\sim x_s$ is evaluated as $F_2\sim N^2TQ$ (see Sect. 4 for more
general results).  These results for the structure functions represent
the unitarity limit for the current--plasma scattering, which in this
case is saturated by the complete absorption of the current --- a genuine
`black disk' limit.

The physical interpretation of these results at small $x$ in terms of
partons in the plasma requires some care: the plasma being infinite, one
needs to take into account the finite duration of the interaction, and
also make a boost to a Lorentz frame where the notion of a parton makes
sense (all the other calculations being done in the plasma rest frame).
But after this is properly into account, it becomes clear that our
results have a natural partonic interpretation, which moreover is
consistent with the corresponding picture for a hadron, as obtained in
Ref. \cite{Hatta:2007he}. Namely, for a given resolution $Q^2$, the
partons exist only at sufficiently small values of $x$, such that
$x\lesssim T/Q$, and are homogeneously distributed in the
three--dimensional phase--space, with occupation numbers of order one.
Equivalently, for a given value of $x$, partons exist only at transverse
momenta smaller than, or equal to, the saturation momentum $Q_s(x)\simeq
T/x$. This value for the saturation momentum is consistent with the
representation of the ${\mathcal N}=4$ SYM plasma as an incoherent
superposition of thermal quasiparticles.

It is finally interesting to notice a similarity between our above
estimate for the plasma saturation momentum and some results in the
literature  \cite{Liu:2006nn,Chernicoff:2006hi} for the {\em screening
length} $L_s(v,T)$ of a heavy quark--antiquark pair moving at velocity
$v$ in the hot ${\mathcal N}=4$ SYM plasma. The screening length is the
maximal separation for which the quark and the antiquark can be still
connected by a string `hanging down' in the radial direction of $AdS_5$.
In Refs. \cite{Liu:2006nn,Chernicoff:2006hi}, one found $L_s(v,T)\simeq
\kappa(1-v^2)^{1/4}/T$ with $\kappa$ a numerical constant (at least for
$v$ close to 1). Now, in the analogy with our DIS problem, the
`quark--antiquark pair' of Refs. \cite{Liu:2006nn,Chernicoff:2006hi}
corresponds to the SYM system emerging from the ${\mathcal R}$--current,
which has a typical transverse extent $1/Q$ and a rapidity $q/Q$. It is
therefore natural to identify our variables $1/Q$ and $q/Q$ with the size
$L$ and the Lorentz gamma factor $\gamma=1/\sqrt{1-v^2}$ of the $q\bar q$
pair, respectively. In particular our saturation momentum $Q_s(x,T$)
should be compared to the inverse screening length $1/L_s(v,T)$. To that
aim, it is preferable to rewrite the result in Refs.
\cite{Liu:2006nn,Chernicoff:2006hi} as $1/L_s^2\sim \gamma T^2$; after
replacing $1/L_s\to Q_s$ and $\gamma\to q/Q_s$, this translates into
$Q_s^3(q,T)\sim qT^2$, which is parametrically the same as our result for
$Q_s$, as alluded to above. It would be interesting to explore this
correspondence in more detail.

\section{General setup and basic equations}
\setcounter{equation}{0}

Following the general strategy with the problem of deep inelastic
scattering, our objective will be to compute the retarded
current--current commutator
  \beq
 R_{\mu\nu}(q)\,=\,i\int \rmd^4x\,\rme^{-iq\cdot x}\,\theta(x_0)\,
 \langle [J_\mu(x), J_\nu(0)]\rangle\,, \label{1}  \eeq
whose imaginary part determines the DIS structure functions. In the
present context, the density $J_\mu(x)$ which enters Eq.~(\ref{1}) refers
to an ${\mathcal R}$--current --- the conserved current associated with a
gauged $U(1)$ subgroup of the $SU(4)$ global ${\mathcal R}$-symmetry
---, and the expectation value is understood as a thermal average, over
the statistical ensemble corresponding to a ${\mathcal N}=4$ SYM plasma
at temperature $T$. The operator $J_\mu(x)$ for the ${\mathcal
R}$--current receives contributions from the fermionic and scalar fields
of the ${\mathcal N}=4$ SYM theory. Accordingly, and following the
example of perturbative QCD, we expect the imaginary part of
$R_{\mu\nu}(q)$ to give us information about the constituents of the
finite--temperature plasma, just as the structure function of the proton
gives information on its (partonic) structure in perturbative QCD.

In the limit where the Yang--Mills coupling $g^2$ is small but the `t
Hooft coupling $\lambda=g^2N$ is large, the AdS/CFT correspondence allows
one to evaluate Eq.~(\ref{1}) in terms of classical supergravity in the
metric of the $AdS_5\times S^5$ black hole. The corresponding metric
reads
 \beq \label{met1} \rmd s^2=\frac{(\pi TR)^2}{u}(-f(u)\rmd t^2+\rmd
 \bm{x}^2)+\frac{R^2}{4u^2f(u)}\rmd u^2 +R^2\rmd \Omega_5^2\,,
 \eeq
where  $T$ is the temperature of the black hole (the same as for the
${\mathcal N}=4$ SYM plasma), $R$ is the common radius of $AdS_5$ and
$S^5$, $t$ and $\bm{x}=(x,y,z)$ are the time and, respectively, spatial
coordinates of the physical Minkowski world, $u$ is the radial coordinate
on $AdS_5$,  $\rmd\Omega^2_5$ is the angular measure on $S^5$, and
$f(u)=1-u^2$. Note that our radial coordinate has been rescaled in such a
way to be dimensionless: in terms of the more standard, dimensionfull,
coordinate $r$, it reads $u\equiv (r_0/r)^2$, with $r_0=\pi R^2 T$.
Hence, in our conventions, the black hole horizon lies at $u=1$ and the
Minkowski boundary at $u=0$.

In order to evaluate Eq.~(\ref{1}), one needs to study the metric
perturbation induced by the $R$--current $J_\mu$ around the background
metric (\ref{met1}). The relevant gravitational wave is a vector field
$A_m(t,\bm{x},u)$ in $AdS_5$, which obeys the classical equations of
motion with given boundary conditions at $u=0$. (Here, $m=\mu$ or $u$ is
the coordinate index on $AdS_5$, with $\mu=0,1,2,3$ referring to a
Minkowski coordinate.) Once the corresponding solution is known, the
tensor $R_{\mu\nu}$ can be extracted from the classical supergravity
action evaluated as a functional of the boundary fields
$A_\mu(t,\bm{x},0)$ (see below). We shall assume the external current to
be weak, so that the metric perturbations be small and the corresponding
equations be linear in $A_m$. Accordingly, we only need the supergravity
action to quadratic order in $A_m$, which reads (see, e.g.,
\cite{Son:2007vk})
   \beq\label{action0} S\,=\,-\frac{N^2}{64\pi^2R}\int \rmd^4x\,
   \rmd u\, \sqrt{-g}\,g^{mp}g^{nq}\,F_{mn}F_{pq}\,, \eeq
where $F_{mn}=\partial_m A_n-\partial_n A_m$, $\partial_m=
\partial/\partial x^m$ with $x^m=(t,\bm{x},u)$, and $g=\det(g_{mn})$.
The classical equations of motion generated by the action (\ref{action0})
are the Maxwell equations in the geometry of the $AdS_5$ black hole. We
shall work in the gauge $A_u=0$ and choose the incoming perturbation as a
plane wave propagating in the $z$ direction: $q^\mu=(\omega,0,0,q)$. Then
we can write
   \beq \label{pw}
   A_\mu(t,\bm{x},u)\,=\,\rme^{-i\omega t+iq z}\,A_\mu(u) \eeq
with the fields $A_\mu(u)$ obeying the following equations (below,
$i=1,2$)
   \beq \varpi A^\prime_0+kfA_3^\prime \,=\,0 \label{11} \\[0.2cm]
   A_i^{\prime\prime}+\frac{f^\prime}{f}A_i^\prime +
   \frac{\varpi^2-k^2f}{uf^2}A_i\,=\,0 \label{ai} \\
   A_0^{\prime\prime}-\frac{1}{uf}(k^2A_0+\varpi k A_3)\,=\,0  \label{13} \eeq
where a prime on a field indicates a $u$--derivative and we have
introduced dimensionless, energy and longitudinal momentum, variables,
defined as
 \beq\label{dimko}
\varpi\equiv \frac{\omega}{2\pi T}\,, \qquad  k\equiv \frac{q}{2\pi T}\,.
 \eeq
Denoting $a(u)\equiv A_0^\prime (u)$, Eqs.~(\ref{11}) and (\ref{13}) can
be combined to give
 \beq
a^{\prime\prime}+\,\frac{(uf)^\prime}{uf}\,a^\prime +
\,\frac{\varpi^2-k^2f}{uf^2}\, a\,=\,0\,, \label{key}
   \eeq
which will be one of our key equations in what follows (the other one
being Eq.~(\ref{ai}) for $A_i$).

The above equations (\ref{11})--(\ref{key}) have already been presented
in the literature (see, e.g., Refs.
\cite{Klebanov:1997kc,Son:2002sd,Herzog:2002pc,Teaney:2006nc,Son:2007vk}),
but in relation with other physical problems, corresponding to physical
regimes very different from ours. These equations must be solved with the
condition that the fields take generic values $A_\mu=A_\mu(u=0)$ at the
$AdS_5$ boundary $u=0$. Then Eq.~(\ref{13}) implies the following
boundary condition for $a(u)$
 \beq \lim_{u\to 0}\big[u a'(u)\big]\,=\,k(kA_0+\varpi A_3)\big|_{u=0}
 \,\equiv\,k^2\mathcal{A}_L(0)\,.
 \label{bc} \eeq
For the solutions to be uniquely specified, an additional boundary
condition is still needed. Following Refs. \cite{Son:2002sd,Son:2007vk},
we shall require the solution to be a purely outgoing wave near the
horizon at $u=1$, where by `outgoing' we mean a wave which is impinging
into the black hole (and thus is departing from the Minkowski boundary).
Physically, this corresponds to the fact that a wave cannot be reflected
by the black hole, but only absorbed. Notice that, in the
zero--temperature case where there is no black hole and the coordinate
$u$ extends to infinity, the corresponding boundary condition is simply
that the fields be regular at $u\to\infty$.

Once the classical solution is known, the next step is to compute the
corresponding, `on--shell', value of the action. Starting with
Eq.~(\ref{action0}) and using the equations of motion to perform the
integration over $u$, it is straightforward to deduce
   \beq S=-\frac{N^2T^2}{16}\int \rmd^4 x
   \left[\left(A_0+\frac{\varpi}{k}A_3\right)a(u)
   -A_i\partial_uA_i(u)\right]_{u=0}\,, \label{actioncl} \eeq
where we have dropped a contribution coming from $u=1$ in accordance with
the prescription in Ref. \cite{Son:2002sd,Herzog:2002pc}. Note that the
appearance of the factor $T^2$ in front of $S$ is merely a consequence of
our definition of the variable $u$ (which scales like $T^2$, so
$\partial_u\sim 1/T^2$). If one returns to the dimensionfull radial
coordinate $r$, then there is no apparent factor $T^2$, and indeed
Eq.~(\ref{actioncl}) has a non--trivial limit as $T\to 0$, corresponding
to the vacuum polarization tensor for the ${\mathcal R}$--current (see
Sect. 3). Given the plane--wave structure in Eq.~(\ref{pw}), the action
density in Eq.~(\ref{actioncl}) is independent of $x^\mu=(t,\bm{x})$, so
it is convenient to separate out the volume of space--time: $S=\int\rmd^4
x\, \mathcal{S}$. From the action density $\mathcal{S}$, the tensor
$R_{\mu\nu}(q)$ is finally obtained as
 \beq R_{\mu\nu}(q)\,=\,\frac{\partial^2 \mathcal{S}}{\partial A_\mu
 \partial A_\nu}\,,
 \label{SR} \eeq
where $A_\mu\equiv A_\mu(u=0)$. Note that the ensuing tensor has mass
dimension two, as it should.

The tensor $R_{\mu\nu}$ can be given the standard tensorial decomposition
(see Appendix A), which shows that there are only two independent scalar
components, $R_1$ and $R_2$, whose imaginary parts determine the two DIS
structure functions, $F_1$ and $F_2$. (The precise definitions are given
in Appendix A.) Since in practice we shall solve second--order
differential equations with {\em real} coefficients, cf. Eqs.~(\ref{ai})
and (\ref{key}), it is interesting to understand how an {\em imaginary}
part in the respective solutions (and hence a non--vanishing contribution
to the DIS structure functions) can arise in the first place. This is
generated via the aforementioned boundary condition near $u= 1$, which
allows for the absorption of the gravitational wave by the black hole.

For a given temperature $T$ of the target plasma, the scalar functions
$R_{i}$, or $F_i$, with $i=1,2$, depend in general upon two kinematical
invariants, that we shall conveniently choose as the virtuality $Q^2$ of
the ${\mathcal R}$--current and the Bjorken $x$ variable. These are
defined as
 \beq\label{Qxdef} Q^2\,\equiv\, q^2-\omega^2\,,\qquad
       {x}\,\equiv\,\frac{Q^2}{-2(q\cdot n)
       T}\,=\,\frac{Q^2}{2\omega T}\,,\eeq
where $n^\mu$ is the four--velocity of the plasma in a generic frame, and
the second expression for $x$ holds in the plasma rest frame, for which
$n^\mu=(1,0,0,0)$. Unless otherwise specified, in what follows we shall
always work in the plasma rest frame. We shall consider the large--$Q^2$
and high--energy kinematics, where $q^2\gg Q^2\gg T^2$ and hence
$\omega\simeq q$. These conditions allow for both small ($x\ll 1$) and
large ($x\sim\order{1}$) values of $x$, but in what follows we shall be
mostly interested in small--$x$ regime where $q\gg Q^2/T$.

To conclude this section, let us present an alternative form for our key
equations, (\ref{ai}) and (\ref{key}), which is more insightful and also
better suited for constructing approximate solutions via WKB techniques.
Via simple manipulations, these equations can be brought into the form of
the (time--independent) Schr\"odinger equation in one spatial dimension,
that is, $\psi^{\prime\prime}-V\psi=0$.

Consider first Eq.~(\ref{key}): when rewritten for the new field
$\psi(u)\equiv \sqrt{u(1-u^2)}\,a(u)$, this takes the form (with
$K^2\equiv k^2-\varpi^2$)
 \beq
 \psi^{\prime\prime}-\frac{1}{u(1-u^2)^2}
  \left[-\frac{1}{4u}(1+6u^2-3u^4)+K^2-k^2u^2\right] \psi\,=\,0\,,
  \label{ad}  \eeq
which is of the Schr\"odinger type, as anticipated. In the interesting
regime at $k^2\gg K^2\gg 1$, the potential $V(u)$ in (\ref{ad}) is well
approximated by
  \beq V=\frac{1}{u(1-u^2)^2}\left[-\frac{1}{4u}+K^2-k^2u^2\right].
   \label{pot} \eeq
This describes a potential barrier, whose shape is illustrated in Fig.1,
and also in Fig. 2a, for three different physical situations,
corresponding to different regimes for the ratio $k/K^3$ : \texttt{(i)}
$k/K^3<{8}/({3\sqrt{3}})$ in Fig.1a, \texttt{(ii)}
$k/K^3={8}/({3\sqrt{3}})$ in Fig.1b, and \texttt{(iii)}
$k/K^3>{8}/({3\sqrt{3}})$ in Fig.1c. As it should be clear from Fig.1b,
the critical value $k/K^3=8/(3\sqrt{3})$ corresponds to the case where
the height of the potential vanishes at its peak. Note that a value of
$\order{1}$ for the ratio $k/K^3$ corresponds to a value $x\sim T/Q\ll 1$
for the Bjorken variable.

We can understand much about the solution to (\ref{ad}) by inspection of
these figures: When $k/K^3<{8}/({3\sqrt{3}})$ (the situation at
intermediate energies), there is a high potential barrier (cf. Fig.1a),
with classical turning points $u_1\simeq 1/(4K^2)$ and $u_2\simeq K/k$.
(Note that $u_1 \ll u_2 \ll 1$ in the interesting regime where $k\gg K\gg
1$, with $k\ll K^3$ though.) We then expect the solution $\psi(u)$ to be
concentrated within the classically allowed region at $u\lesssim 1/K^2$.
Moreover, the DIS structure functions are expected to be extremely small
in this case, since an imaginary part in the classical solution can
develop only via tunneling through the high potential barrier.

On the other hand when $k/K^3>{8}/(3\sqrt{3})$ (the high--energy case,
cf. Fig.1c), there is no potential barrier any longer, so the
gravitational wave can easily flow into the black hole and thus get
absorbed by the latter. We then expect a large imaginary part to
$R_{\mu\nu}$.

\begin{figure}
\centerline{
\includegraphics[width=6cm]{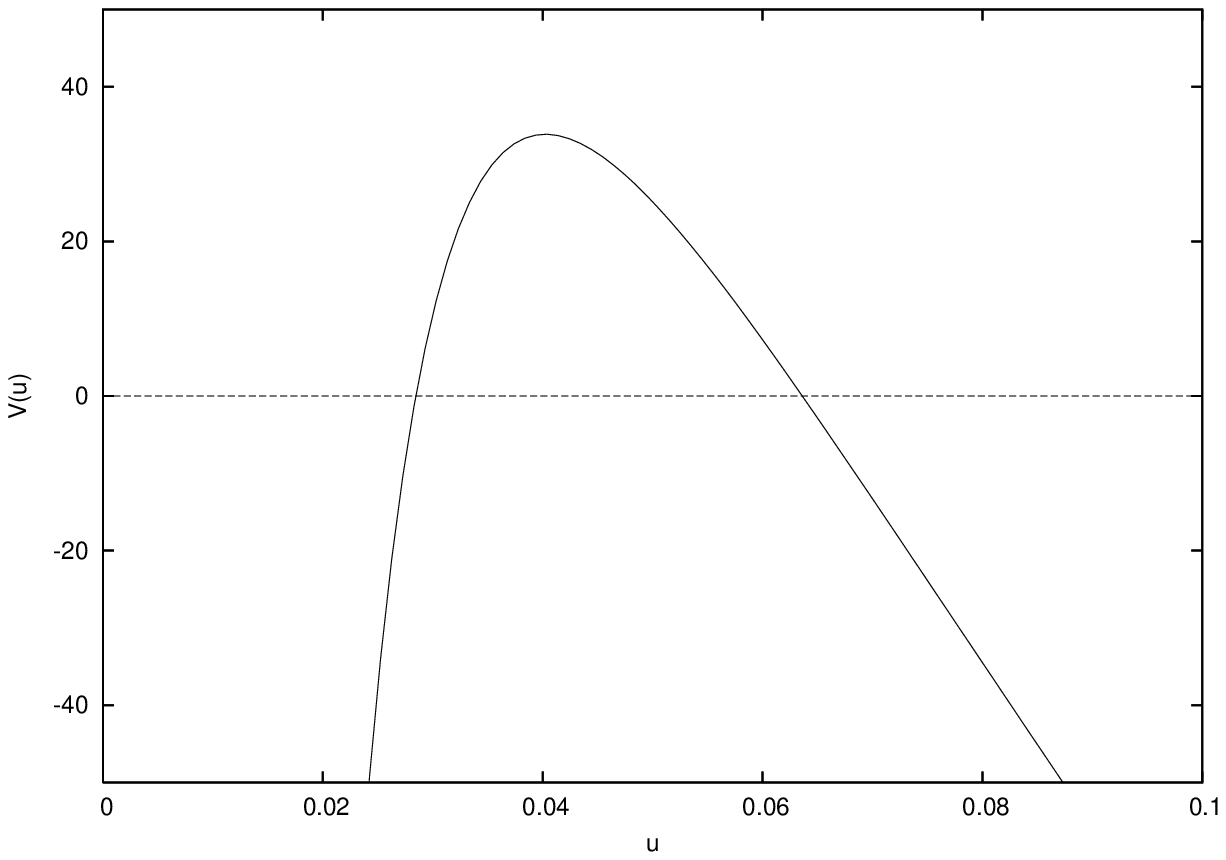}
\includegraphics[width=6cm]{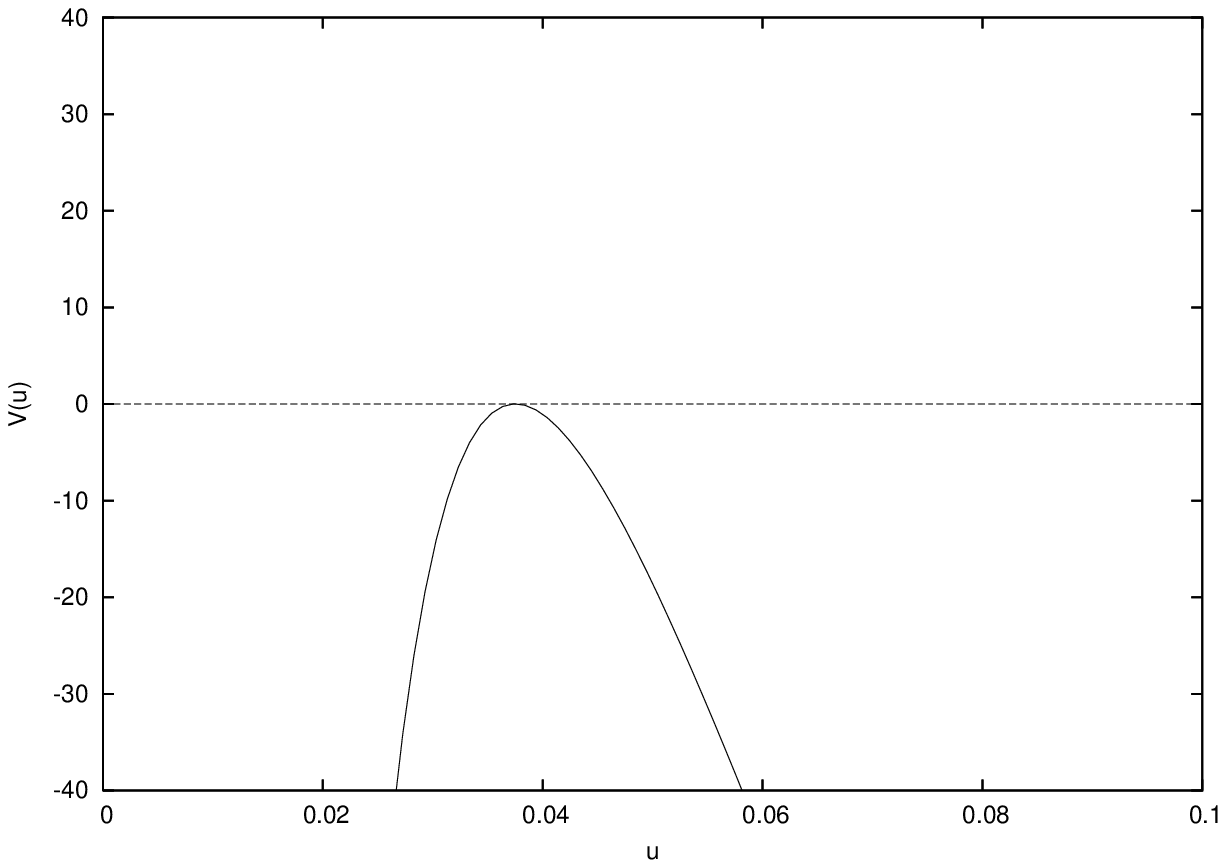}
\includegraphics[width=6cm]{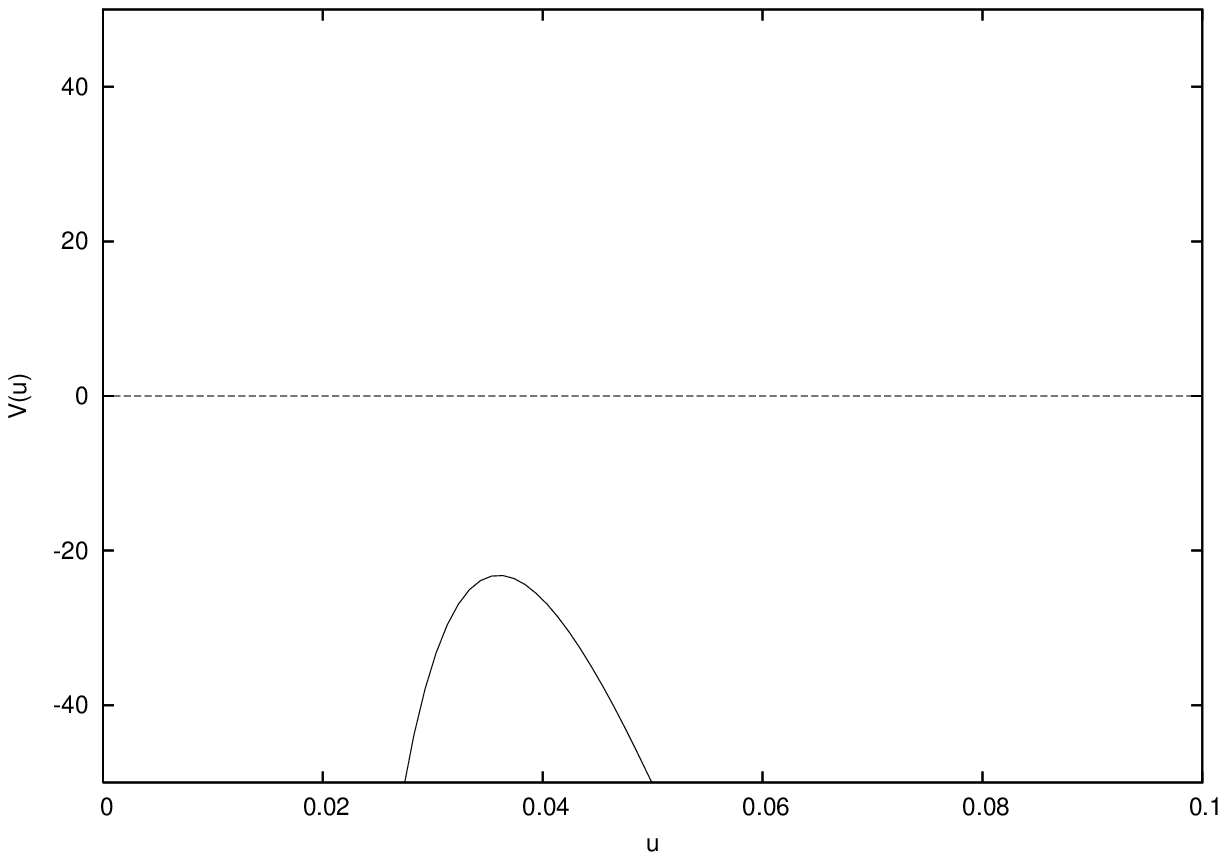}}
\centerline{\sf (a)\hspace*{6cm}(b)\hspace*{6cm}(c)}
{\caption\sl The potential $V(u)$ in Eq.~(\ref{pot}) for three values
of the ratio $k/K^3$: (a) $k/K^3<{8}/({3\sqrt{3}})$, (b)
$k/K^3={8}/({3\sqrt{3}})$, and (c) $k/K^3>{8}/({3\sqrt{3}})$.
For the figures to look more suggestive, all the chosen values for $k/K^3$
are relatively close to the critical value ${8}/({3\sqrt{3}})$.}
 \label{Fig1}
\end{figure}

\begin{figure}
\centerline{
\includegraphics[width=8.5cm]{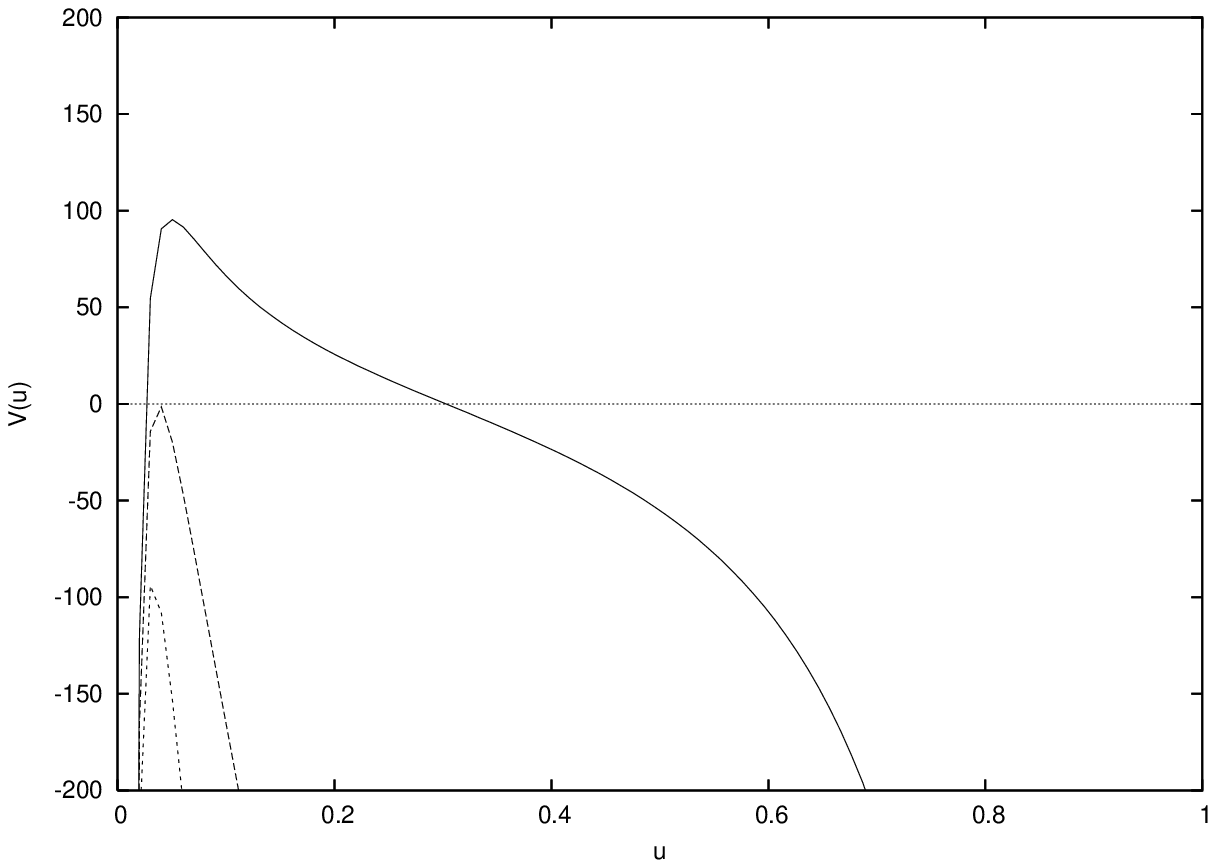}
\includegraphics[width=8.5cm]{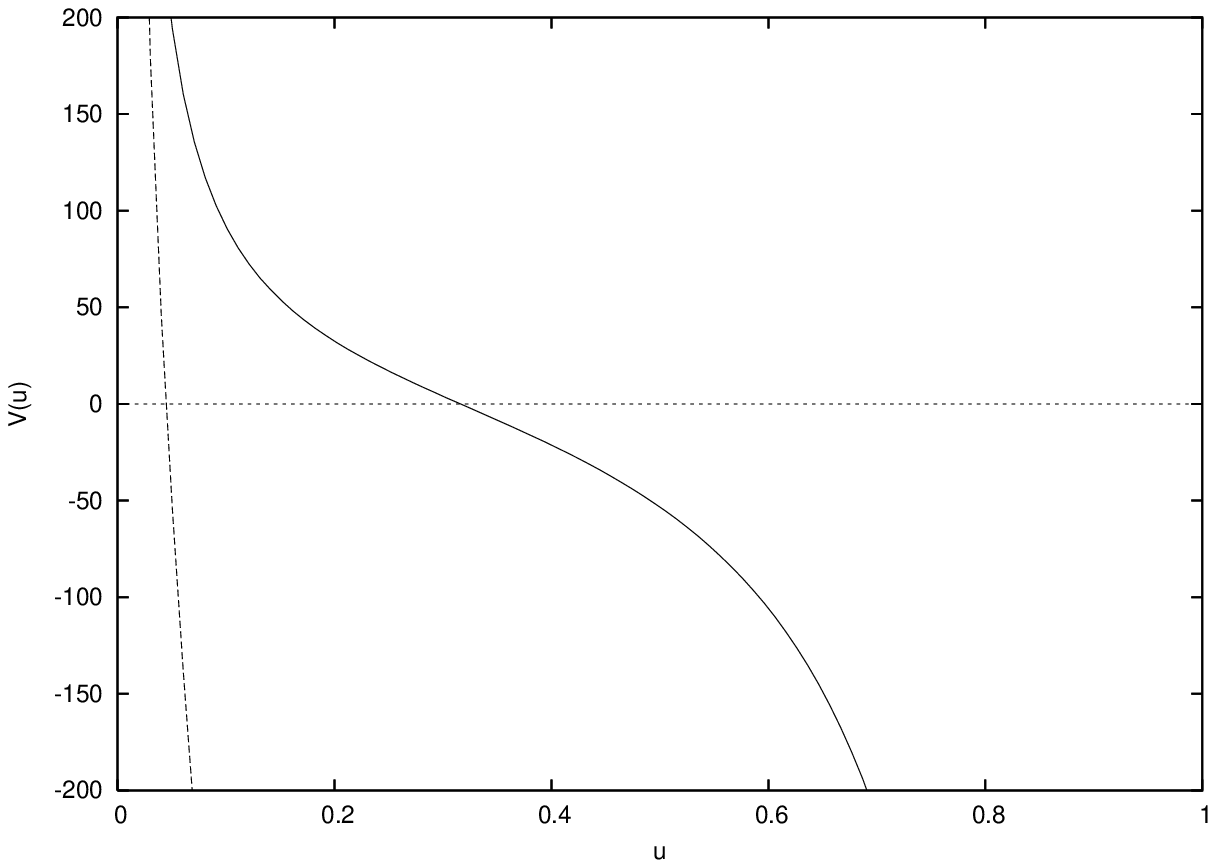}}
\centerline{\sf (a)\hspace*{9cm}(b)}
{\caption\sl The potentials $V(u)$ corresponding to longitudinal
waves, cf. Eq.~(\ref{pot}) (left) and, respectively, transverse waves,
cf. Eq.~(\ref{tpot}) (right), are represented for several values of
the ratio $k/K^3$, corresponding to physical regimes well
separated from each other.}
 \label{Fig2}
\end{figure}

Similar conclusions apply to the transverse modes $A_i$ as well, although
the corresponding argument is slightly more involved, and perhaps less
intuitive. With the substitution $\phi(u)\equiv \sqrt{(1-u^2)}\,A_i(u)$
(for either $i=1$ or $i=2$, the respective equations being identical),
Eq.~(\ref{ai}) takes the Schr\"odinger--like form
 \beq
 \phi^{\prime\prime}-\frac{1}{u(1-u^2)^2}
  \left[K^2-k^2u^2-u\right] \phi\,=\,0\,,
  \label{phi}  \eeq
where for the present purposes the potential can be approximated by
 \beq V=\frac{1}{u(1-u^2)^2}\left[K^2-k^2u^2\right].
   \label{tpot} \eeq
This potential is illustrated in Fig. 2b, for two values of the ratio
$k/K^3$. As manifest in these figures, the potential barrier is now
concentrated near the boundary at $u=0$, within a distance  $u\lesssim
1/K^2$, whereas the classically allowed region (i.e., the region where
$V(u) \le 0$) starts at $u=K/k$. With increasing energy, the barrier does
not disappear anymore, rather it gets squeezed towards $u=0$, in such a
way that its effects become smaller and smaller. At low energy, the wave
can penetrate into the bulk only up to a small distance $u\sim 1/K^2$
away from the boundary. But when the energy is so high that $k/K^3 \sim
\order{1}$, the penetration distance $\sim 1/K^2$ becomes of the same
order as the classical turning point at $K/k$, and then the wave can
freely escape in the allowed region at $u > K/k$, and thus get absorbed
by the black hole.

These simple observations will be confirmed and substantiated by the
subsequent analysis in this paper.

\section{Low energy: the multiple scattering series}
\setcounter{equation}{0}

In this section we shall consider the low--energy regime at $k/K^3\ll 1$,
cf. Fig. 1a, where the effects of the term proportional to $k^2$ in the
potential $V(u)$ (in either Eq.~(\ref{pot}), or (\ref{tpot})) can be
treated in perturbation theory. Note that, in terms of our original
variables $\omega$ and $q$, cf. Eq.~(\ref{dimko}), the condition $k\ll
K^3$ amounts to $qT^2\ll Q^3$. Hence, for a fixed virtuality $Q^2$, the
`low--energy' regime can be also understood as a {\em low--temperature}
one, $T\ll(Q^3/q)^{1/2}$, and the perturbative expansion that we shall
shortly construct can be alternatively viewed as a multiple scattering
series, or a low--temperature expansion.

Clearly, even when $k\ll K^3$, this perturbative expansion cannot work
for arbitrary values of $u$\,: when $u\gtrsim K/k$, the energy--enhanced
term $\propto k^2$ in the potential becomes the {\em dominant} term
there, which is responsible for the existence of the classically allowed
region at $u\ge K/k$. Thus, not surprisingly, the perturbative treatment
of the finite--energy/temperature effects cannot account for the
contributions due to tunneling, which are genuinely non--perturbative and
will be estimated in Appendix B within the WKB approximation. But if one
leaves these contributions aside (they are exponentially suppressed
anyway; see Appendix B), then perturbation theory should work reasonably
well in the small--$u$ region at $u\lesssim 1/K^2$, which is the relevant
region for computing the ${\mathcal R}$--current correlator, cf.
Eqs.~(\ref{actioncl})--(\ref{SR}).

The main result that we shall arrive at in this section could be
characterized as {\em negative}\,: we shall find that for $k \ll K^3$ the
DIS structure functions are strictly zero when computed to all orders in
the multiple scattering (or `twist') expansion. But the subsequent
analysis is still interesting in that it provides the twist expansion for
the real part of $R_{\mu\nu}$. In particular, from the leading term in
this expansion (the single scattering approximation), we shall be able to
deduce a couple of energy--momentum sum rules which will be very useful
later on.

In the interesting region at $u\lesssim 1/K^2\ll 1$, our key equations
(\ref{ai}) and (\ref{key}) simplify to
 \beq
    A_i^{\prime\prime} -
   \frac{K^2}{u}\,A_i\,=\,-k^2uA_i\,, \label{aipert}
 \eeq
and, respectively,
 \beq
 a^{\prime\prime}+\frac{1}{u}\,a^\prime-\frac{K^2}{u}\,a\,=\,-k^2ua\,.
 \label{keypert} \eeq
In writing these equations, we have separated the terms $\propto k^2$ in
the r.h.s., anticipating that they are going to be treated as `small
perturbations'. For consistency with the present approximations, which
ignore the phenomenon of tunneling, the above equations must be solved
with the condition that the fields vanish as $u\to\infty$. (This would be
the correct boundary condition in the zero--temperature limit $T\to 0$,
and it remains the appropriate boundary condition for a perturbative
treatment of the finite--temperature effects.)

Consider first Eq.~(\ref{keypert}); after a change of variable
$\zeta\equiv 2K\sqrt{u}$, this becomes
 \beq
  \left(\frac{\rmd^2}{\rmd\zeta^2}+\frac{1}{\zeta}
  \frac{\rmd}{\rmd\zeta}-1\right)\,a(\zeta)\,=\,
  -\frac{k^2\zeta^4}{16K^6}\,a(\zeta)\,. \label{apert} \eeq
The zero--temperature limit\footnote{The limit $T\to 0$ of the present
equations may look tricky since we have defined the dimensionless
variables in Eq.~(\ref{dimko}) by dividing though $T$. However, in the
zero--temperature case, one can view $T$ in Eq.~(\ref{dimko}) as an
arbitrary reference scale, introduced in order to define dimensionless
variables. This scale cancels out in the final results for the current
correlator at $T=0$, as one can check on the examples of Eqs.~(\ref{S0})
and (\ref{R0}) below.} of this equation, that is,
 \beq \left(\frac{\rmd^2}{\rmd\zeta^2}+\frac{1}{\zeta}
 \frac{\rmd}{\rmd\zeta}-1\right)a^{(0)}(\zeta)\,=\,0\,,\label{a0eq}\eeq
describes the (longitudinal) metric perturbation induced by the
${\mathcal R}$--current in $AdS_5$ in the absence of the black hole (the
supergravity dual of an ${\mathcal R}$--current propagating through the
gauge theory vacuum). The general solution to (\ref{a0eq}) is a linear
combination of the modified Bessel functions $\mathrm{K}_0$ and
$\mathrm{I}_0$. The coefficient of $\mathrm{I}_0$ is set to zero by the
condition of regularity as $\zeta\to\infty$, while that of $\mathrm{K}_0$
is fixed by the boundary condition at $\zeta=0$, cf. Eq.~(\ref{bc}). One
thus finds
  \beq a^{(0)}(\zeta)
  =-2k^2\mathcal{A}_L(0)\,\mathrm{K}_0(\zeta)\,. \label{a0sol} \eeq
The general equation (\ref{keypert}) can be given a formal solution via
Green's function techniques :
 \beq\label{asol} a(\zeta)\,=\,a^{(0)}(\zeta)\,+\int_0^\infty \rmd
 \zeta^\prime \,G(\zeta,\zeta^\prime)\,\left(\frac{-k^2\zeta^{\prime
 4}}{16K^6}\right) \,a(\zeta^\prime)\,, \eeq
with the Green's function $G(\zeta,\zeta^\prime)$ obeying
 \beq
 \left(\frac{\rmd^2}{\rmd\zeta^2}+\frac{1}{\zeta}
 \frac{\rmd}{\rmd\zeta}-1\right)G(\zeta,\zeta^\prime)=\delta(\zeta-\zeta^\prime)\,,
 \label{green}  \eeq
together with the following boundary conditions
 \beq G(\zeta,\zeta^\prime)&\ \to\ & 0 \quad\mbox{as}\quad \zeta\to
 \infty,\nn
 \zeta \frac{\rmd}{\rmd\zeta} \,G(\zeta,\zeta^\prime)&\ \to\ & 0
  \quad\mbox{as}\quad \zeta\to 0. \eeq
It is easily checked that the corresponding solution reads
 \beq\label{Green} G(\zeta,\zeta^\prime)\,=\,-\zeta^\prime\,
 \big\{\mathrm{K}_0(\zeta)\mathrm{I}_0(\zeta^\prime)
  \Theta(\zeta-\zeta^\prime)+\mathrm{K}_0(\zeta^\prime)\mathrm{I}_0(\zeta)
  \Theta(\zeta^\prime-\zeta)\big\}\,. \eeq
The `solution' (\ref{asol}) is truly an integral equation for $a(\zeta)$,
which generates the multiple scattering series through iterations ---
here, for the longitudinal wave.

Consider similarly the transverse sector. By replacing $A_i=A_i(0)\zeta
h(\zeta)$, with $\zeta= 2K\sqrt{u}$, within Eq.~(\ref{aipert}), one finds
   \beq
   \left(\frac{\rmd^2}{\rmd\zeta^2}+\frac{1}{\zeta}
   \frac{\rmd}{\rmd\zeta}-1-\frac{1}{\zeta^2}\right)h
   =-\frac{k^2\zeta^4}{16K^6}\,h\,. \label{46} \eeq
The zero--temperature version of this equation is solved by $h^{(0)}
=\mathrm{K}_1(\zeta)$, which obeys $\zeta h^{(0)}(\zeta)\to 1$ as
$\zeta\to 0$, as it should. (The other solution $\mathrm{I}_1(\zeta)$ is
rejected by the condition of regularity at infinity.) The general
equation (\ref{46}) can then be rewritten as an integral equation similar
to Eq.~(\ref{asol}) with $a^{(0)}\to h^{(0)}$ and the following Green's
function
 \beq
  G(\zeta,\zeta^\prime)\,=\,-\zeta^\prime\,\big\{\mathrm{K}_1(\zeta)
  \mathrm{I}_1(\zeta^\prime)\Theta(\zeta-\zeta^\prime)+\mathrm{I}_1(\zeta)
  \mathrm{K}_1(\zeta^\prime)\Theta(\zeta^\prime-\zeta)\big\}. \eeq

As a simple application of the previous results, let us now compute the
first two terms in the low--temperature expansion of the current--current
correlator (\ref{1}) --- that is, its zero--temperature piece
$R^{(0)}_{\mu\nu}$, which represents the vacuum polarization tensor of
the ${\mathcal R}$--current, and the first temperature--dependent
contribution $R^{(1)}_{\mu\nu}$, which describes the scattering between
the ${\mathcal R}$--current and the ${\mathcal N}=4$ SYM plasma in the
single--scattering, or `leading twist', approximation. To that aim, we
need the first two iterations in the above integral equations for $a(u)$
and $A_i(u)$, evaluated near $u=0$ (cf.
Eqs.~(\ref{actioncl})--(\ref{SR})).

To the order of interest, we can write $a(u)=a^{(0)}(u)+a^{(1)}(u)$,
where $a^{(0)}(u)$ has a logarithmic singularity as $u\to 0$, as expected
according to Eq.~(\ref{bc}),
 \beq a^{(0)}(u)\,=\,
  k^2\mathcal{A}_L(0)\,
  \big(\ln K^2+\ln u +2\gamma + \order{u}\big)\, \label{a0} \eeq
($\gamma=0.577...$ is Euler's constant), while $a^{(1)}(0)$ is finite and
equal to
 \beq a^{(1)}(0)\,=\,\mathcal{A}_L(0)\,\frac{-2k^4}{16K^6}\,
   \int \rmd \zeta \,\zeta^5 \mathrm{K}_0^2(\zeta) =
   -\,\frac{2k^4}{15K^6}\,
   \mathcal{A}_L(0)\,. \label{a1}
   \eeq
As for the transverse fields $A_i(u)$, these are needed up to linear
order in $u$, i.e., to quadratic order in $\zeta$\,; one finds
$A_i(u)=A_i^{(0)}(u)+A_i^{(1)}(u)$, with
   \beq A_i^{(0)}(u)&\,\simeq\,&A_i(0)\Big\{1+uK^2
 \big(\ln K^2 -1 +\ln u +2\gamma\big)\Big\}\label{Ai0}\\[0.2cm]
   A_i^{(1)}(u)&\,\simeq\,& A_i(0)\,\frac{u k^2}{5K^4}\,.\label{Ai1}\eeq
When the $T=0$ fields in (\ref{a0}) and (\ref{Ai0}) are used to evaluate
the vacuum action $S^{(0)}$, cf. Eq.~(\ref{actioncl}), the result
exhibits a logarithmic divergence coming from the limit $u\to 0$. As
usual in the AdS/CFT context, this singularity is interpreted as a
ultraviolet divergence in the dual gauge theory, to be removed via
renormalization. To that aim, it is important to return to the original
variables $r$, $\omega$, $q$, and $Q^2=q^2-\omega^2$, to make it clear
that the UV `counterterms' are indeed temperature--independent. Then, the
relevant terms in the action are
 \beq
 \ln K^2  +\ln u \,= \,\ln\frac{Q^2}{4\pi^2T^2}+
 \ln\frac{\pi^2 R^4T^2}{r^2} \,= \, \ln\frac{Q^2}{\Lambda^2}+
 \ln\frac{R^4 \Lambda^2}{4r^2}\,,\eeq
where the $T$--dependence has disappeared, as anticipated, and $\Lambda$
plays the role of the substraction scale on the gauge theory side. For
convenience, we renormalize by dropping the last term in the above
equation together with the finite term $2\gamma$. We thus obtain
$S^{(0)}=\int\rmd^4 x\, \mathcal{S}^{(0)}$, with
    \beq
    \mathcal{S}^{(0)}=-\frac{N^2}{64\pi^2}\ln\frac{Q^2}{\Lambda^2}
    \,\big[(qA_0+ \omega A_3)^2-
     Q^2\calbfA_T\cdot \calbfA_T\,\big]_{u=0}\,, \label{S0}
     \eeq
where we have introduced the transverse vector notation $\calbfA_T \equiv
(A_1,A_2)$. From this expression, one can immediately deduce the vacuum
polarization tensor (with $\eta_{\mu\nu}=(-1,1,1,1)$) :
 \beq\label{R0} R_{\mu\nu}^{(0)}(q)=
  \left(\eta_{\mu\nu}-\frac{q_\mu q_\nu}{Q^2} \right)R_1^{(0)}(Q^2)\,\qquad
 \mbox{with}\qquad R_1^{(0)}(Q^2)
 =\frac{N^2Q^2}{32\pi^2}\ln\frac{Q^2}{\Lambda^2}\,.\eeq
This is transverse, as required by current conservation, and moreover it
has exactly the same expression as in zeroth--order (one loop)
perturbation theory. This `non--renormalization' property of the
${\mathcal R}$--current polarization tensor has been already observed in
the literature, and proven to be a consequence of supersymmetry
\cite{Anselmi:1997am}.

Similarly, by using the finite--$T$ contributions to the fields,
Eqs.~(\ref{a1}) and (\ref{Ai1}), one can compute the respective
contribution to the on--shell action, $S^{(1)}=\int\rmd^4 x\,
\mathcal{S}^{(1)}$, which is manifestly ultraviolet--finite :
   \beq
     \mathcal{S}^{(1)}=\frac{N^2\pi^2T^4}{30}\,\frac{q^2}{Q^6}\,
     \left[(q A_0+\omega A_3)^2+\frac{3}{2}\,Q^2\calbfA_T^2
    \right]_{u=0}\,, \label{S1} \eeq
From (\ref{S1}) one can determine the single--scattering, or
low--temperature, part of the tensor $R_{\mu\nu}$. This has the structure
exhibited in Eq.~(\ref{A.1}) with the following, `leading--twist',
expressions for the scalar components $R_1$ and $R_2$ :
 \beq R_1^{(1)}=\frac{N^2\pi^2T^2}{40x^2}\,,\qquad
 R_2^{(1)}=\frac{N^2\pi^2T^4}{6Q^2}\,, \label{R12LT}
  \eeq
which are both {\em real} : as anticipated at the beginning of the
section, the DIS structure functions vanish in the leading--twist
approximation, and in fact to {\em all} orders in the twist expansion ---
indeed, all the terms generated by iterating the integral equation
(\ref{asol}) for the longitudinal field, or the corresponding equation
for the transverse fields, are obviously real.

Note the $1/x^2$ behavior of $R_1^{(1)}$, which is the hallmark of the
graviton exchange and reflects the fact that the contributions to
$R_{\mu\nu}$ computed in Eq.~(\ref{R12LT}) come from the twist--two and
spin--two operator $T_{\mu\nu}$ (the energy--momentum tensor) in the
operator product expansion (OPE) of the current--current correlator
(\ref{1}). This is the only leading--twist operator which survives in the
OPE at strong coupling, since the other twist--two operators with spins
$j>2$ acquire large anomalous dimensions $\sim \lambda^{1/4}\to \infty$.
It is quite remarkable that the OPE coefficients of $T_{\mu\nu}$ that we
have (indirectly) computed at strong coupling are exactly the same as the
corresponding coefficients at weak coupling, as we shall demonstrate via
an explicit zeroth--order calculation in Appendix C. This
non--renormalization is yet another manifestation of the high degree of
symmetry of the ${\mathcal N}=4$ SYM theory (see also Ref.
\cite{Arutyunov:2000ku}).

Similarly, the multiple scattering series previously discussed can be
interpreted as the exchange of arbitrarily many gravitons. One simple way
of understanding the lack of an imaginary part in these multiple graviton
exchanges is to note that the gravitons carry no four--dimensional
space--time momentum, as reflected in the fact that the metric only
depends upon the radial variable $u$ in $AdS_5$. Hence, because of
energy--momentum conservation, the graviton exchanges cannot create
on--shell final states, which would be the source for inelasticity.

We thus conclude that in the intermediate energy/low temperature regime
at $k\ll K^3$ (or, equivalently, at relatively large values $x\gg T/Q$
for the Bjorken variable), the only non--trivial contributions to the DIS
structure functions $F_i\propto {\rm Im}\,R_i$ arise via tunneling
through the potential, and thus are necessarily small. In Appendix B,
these contributions will be estimated in the WKB approximation as
$F_i\sim \exp\{-c(K^3/k)^{1/2}\}$, where the prefactor $c$ is a number of
$\order{1}$. This estimate confirms that ${\rm Im}\,R_i$ remains
extremely small so long as $k\ll K^3$. We thus draw the rather striking
conclusion that the strongly--coupled plasma has essentially no
point--like constituents at $x$ larger than $x_s\sim T/Q$.

Finally, let us mention an interesting consequence of the leading--twist
results in Eq.~(\ref{R12LT}), which will be very useful in what follows.
Introducing the variable $z\equiv 1/x$ and assuming standard analytic
properties for the current--current correlator in the complex $z$ plane,
one can relate the behaviour of $R_i(z)$ near $z=0$, where
Eq.~(\ref{R12LT}) applies, to the integral of the DIS structure function
$F_i\propto {\rm Im}\,R_i$ along the cuts on the real axis in the
physical region at $|z|>1$. One thus obtain the following sum--rules (see
Appendix A for details)
  \beq \mathcal{E}=18T^2\int_0^1 \rmd x F_2(x,Q^2),
 \label{sum1} \\ \mathcal{E}=45T^2\int_0^1
   \rmd xF_L(x,Q^2), \label{sumL}\eeq
where $F_1$ and $F_2$ are defined in (\ref{F1})--(\ref{F2}),
$F_L=F_2-2xF_1$ is the longitudinal structure function, and
 \beq\label{Theta}
 \mathcal{E}=\frac{3\pi^2N^2T^4}{8} \eeq
is the energy density of the ${\mathcal N}=4$ SYM plasma in the strong
coupling limit: $\mathcal{E}=\Theta_{00}$, with $\Theta_{\mu\nu} $ the
energy--momentum tensor of the plasma, cf. Eq.~(\ref{theta}). The
appearance of the energy density in the l.h.s.'s of equations
(\ref{sum1}) and (\ref{sumL}) is in fact natural: as we shall further
explain in Sect. 5, the integrals in their r.h.s.'s are proportional to
the energy density carried by the plasma constituents, as probed in DIS
with a resolution scale $Q^2$; this should be the same as the total
energy density in the plasma, and in particular be independent of $Q^2$
--- which is precisely the content of Eqs.~(\ref{sum1})--(\ref{sumL}).

But the previous results in this section also show that the relatively
large values of $x$, such that $x > T/Q$, give only tiny contributions to
the structure functions, which die away exponentially at large $Q^2$ and
hence cannot ensure the fulfillment of the sum--rules. Therefore, the
only way for these sum--rules to be satisfied is that the integrals in
their r.h.s.'s be saturated by contributions from `partons' at smaller
values of $x\lesssim T/Q$. This corresponds to the `high--energy'
situation in Fig. 1c, to the analysis of which we now turn.


\section{High energy: deep inelastic scattering}
\setcounter{equation}{0}

In this section, we shall consider the high--energy ($k> K^3$), or
small--$x$ ($x<T/Q$), regime, which is the most interesting regime for
our present analysis, since this is where the deeply inelastic scattering
truly occurs. In this regime, the potential barrier becomes ineffective
--- it has either completely disappeared (in the longitudinal sector, cf.
Figs. 1c or 2a), or become so narrow that it gives no significant
attenuation (in the transverse sector, cf. Fig. 2b) ---, and then the
gravitational waves induced by the ${\mathcal R}$--current can propagate
towards large values of $u\sim\order{1}$, until they reach the black hole
horizon at $u=1$ and thus get absorbed. As explained in Sect. 2, this
absorption manifests itself via imaginary parts in the classical
solutions, that we shall first compute, and from which we shall then
deduce the DIS structure functions.

By lack of exact solutions to the wave equations (\ref{ai}) and
(\ref{key}), we shall consider approximations which are valid for very
high energies, such that $k\gg K^3$, but which cannot capture the
transition from quasi--elastic to deeply--inelastic scattering, which
takes place around $k\sim K^3$. In Appendix D, we shall construct
approximate solutions valid for generic values of $u$, by performing
piecewise approximations (in particular, the WKB approximation) and then
matching the intermediate solutions with each other. Here, however, we
shall use a simpler strategy to calculate the classical action
(\ref{actioncl}). To appreciate this strategy, let us first recall what
was the main difficulty with this calculation: although the action
involves the classical solution near $u=0$ alone, as manifest on
Eq.~(\ref{actioncl}), this solution is generally sensitive to the
dynamics at large $u\sim\order{1}$, via the `outgoing wave' boundary
condition that one has to impose near the horizon. The important
simplification that appears at high energy is that this boundary
condition can now be imposed already at {\em relatively small} values
$u\ll 1$, where the general equations reduce to simpler ones, that can be
solved exactly. Indeed, in the absence of any potential barrier, there is
no mechanism to generate reflected waves at intermediate values of $u<1$;
hence, an incoming wave cannot be tolerated in the solution not even at
$u\ll 1$, since it would necessarily describe reflection off the black
hole. This argument will be confirmed by the more general construction in
Appendix D, which will provide the same small--$u$ solutions as obtained
below in this section.

Note an additional, important, simplification which occurs at high
energy: when $k\gg K^3$, the term in the potential involving the
virtuality $K^2$ of the current becomes negligible as compared to the
other terms there, for all the relevant values of $u$ (for both
longitudinal and transverse modes). This means, in particular, that our
subsequent discussion also applies to a {\em time--like} ($K^2\equiv
k^2-\varpi^2< 0$) current, provided its energy is high enough ($k\gg
|K|^3$).

Indeed, consider the longitudinal sector first. When increasing $u$ from
$u=0$, the last term $\propto k^2u^2$ in the potential (\ref{pot})
becomes comparable to the first term $\propto 1/4u$ already at the very
small value $u_0= 1/(4k^2)^{1/3}$, at which the term $\propto K^2$ is
still negligible. Hence, the latter is never relevant, as anticipated. In
particular, for $u\ll 1$, the potential simplifies to
 \beq\label{Vhigh} V\simeq -\frac{1}{u^2}
 \left[\frac{1}{4}+k^2u^3\right] \qquad\mbox{for}\qquad
   u\ll 1 \,, \eeq
which has a peak at $u\sim u_0$, cf. Fig.1c. By performing the
corresponding approximations on Eq.~(\ref{key}), this equation becomes
 \beq
 a^{\prime\prime}+\frac{1}{u}\,a^\prime+k^2ua\,=\,0\,,
 \eeq
which can be easily solved: after changing variable according to
$\xi\equiv \frac{2}{3}ku^{3/2}$, we obtain
 \beq
 \left(\frac{\rmd^2}{\rmd\xi^2}+\frac{1}{\xi}\frac{\rmd}
 {\rmd\xi} +1\right)a(\xi)=0
 \eeq
which has the general solution
 \beq a(\xi)=c_1\mathrm{J}_0(\xi)+c_2\mathrm{N}_0(\xi)\,, \label{57}\eeq
where $\mathrm{J}_0$ and $\mathrm{N}_0$ are the usual Bessel and Neumann
functions. Recalling the behaviour of these functions near $\xi=0$, one
sees that the boundary condition (\ref{bc}) fixes the coefficient $c_2$,
\beq\label{c2}
 c_2=\frac{\pi k^2}{3}\mathcal{A}_L(0)\,,\eeq
but has no consequence for $c_1$. To also determine the latter, we shall
impose, as announced, the outgoing--wave boundary condition at
sufficiently large values of $u$. Note that, although $u$ is small, $u\ll
1$, the argument $\xi$ of the Bessel functions becomes large, $\xi\gg 1$,
for all the values of $u$ far beyond the peak of the potential: $u\gg
u_0\sim 1/k^{2/3}$. In that region, one can use the asymptotic
expressions for the Bessel functions, that is, $\mathrm{J}_0(\xi)\simeq
\sqrt{2/\pi\xi}\,\cos(\xi-\pi/4)$ and $\mathrm{N}_0(\xi)\simeq
\sqrt{2/\pi\xi}\,\sin(\xi-\pi/4)$. If one also remembers the exponential
factor yielding the time--dependence, cf. Eq.~(\ref{pw}), it becomes
clear that a purely outgoing--wave solution $a(t,\xi)\propto
\rme^{-i(\omega t-\xi)}$ is obtained by choosing
  \beq\label{c12}
  c_1=-ic_2\,. \eeq
One thus obtains the following expression for the longitudinal solution
in the region $u\ll 1$
 \beq a(u)\,\simeq\,-i\,\frac{\pi k^2}{3}\mathcal{A}_L(0)\,\mathrm{H}_0^{(1)}
   \left(\frac{2}{3}ku^{3/2}\right)\qquad\mbox{for}\qquad
   u\ll 1\,, \label{71} \eeq
where $\mathrm{H}_0^{(1)}=\mathrm{J}_0+i\mathrm{N}_0$ is a Hankel
function.

A similar discussion applies to the transverse waves, which satisfy
Eq.~(\ref{ai}), or (\ref{phi}). In the high--energy regime at $k\gg K^3$,
one can neglect the effects of the extremely narrow potential barrier
located at $0<u<K/k$. Indeed, the width $K/k$ of the barrier is much
smaller then the distance $\sim 1/K^2$ over which the solution near $u=0$
would start to significantly differ from its boundary value at $u=0$. In
the small--$u$ region at $K/k \lesssim u\ll 1$, the potential
(\ref{tpot}) reduces to $V\simeq-k^2u$ and then both Eq.~(\ref{ai}) and
Eq.~(\ref{phi}) reduce to
  \beq A_i^{\prime\prime}+k^2uA_i=0\,.
  \label{72}
  \eeq
This is an Airy equation whose general solution can be written as a
linear combination of ${\rm Ai}(-uk^{\frac{2}{3}})$ and ${\rm
Bi}(-uk^{\frac{2}{3}})$ or, equivalently \cite{ab}, in terms of the
Bessel functions of argument $\nu=1/3$. We choose this latter
representation, for more symmetry with the previous discussion; we thus
write (with $\xi= \frac{2}{3}u^{3/2}k$, as before)
 \beq\label{Aismall} A_i(\xi)=\xi^{\frac{1}{3}}
  \big[c_1\mathrm{J}_{{1}/{3}}(\xi)+c_2\mathrm{N}_{{1}/{3}}(\xi)
  \big]\,, \eeq
where $c_2$ is determined from the value of $A_i$ at $u=0$ and we again
choose $c_1 = -ic_2$, in order for the solution to become a purely
outgoing wave when $u\gg  1/k^{2/3}$. One finally gets the following
result at small $u$ :
  \beq A_i(u)\,\simeq\,  A_i(0)\,\frac{i\pi}{\Gamma(1/3)}
  \left(\frac{k}{3}\right)^{{1}/{3}}
\sqrt{u}\ \mathrm{H}_{1/3}^{(1)}
  \left(\frac{2}{3}ku^{3/2}\right)\qquad\mbox{for}\qquad
   u\ll 1\,, \label{81}  \eeq
which now features the Hankel function
$\mathrm{H}_{1/3}^{(1)}=\mathrm{J}_{1/3}+ i\mathrm{N}_{1/3}$.

By putting together the previous results (\ref{71}) and (\ref{81}), one
can evaluate the on--shell action according to (\ref{actioncl}); this
gives $S=\int\rmd^4 x\, \mathcal{S}$ with
  \beq\label{She} \mathcal{S}\,=\,-\frac{N^2T^2}{48}
  \left[k^2\mathcal{A}_L^2(0)
  \left(2\Big(\gamma+\ln\frac{k}{3}\Big)-i\pi\right)\,+
  \,\frac{9\pi}
  {\Gamma^2(1/3)}
  \left(\frac{k}{3}\right)^{2/3}
  \left(\frac{1}{\sqrt{3}}-i\right)
  \calbfA_T^2(0) \right]\,. \nn
  \eeq
Note the emergence of the imaginary part in the action (\ref{She}), which
has the right sign (${\rm Im}\,\mathcal{S} >0$) to describe dissipation,
i.e., to yield positive contributions to the DIS structure functions. A
simple calculation using Eqs.~(\ref{SR}), (\ref{A.1}), (\ref{F1}) and
(\ref{F2}), finally leads to the following expressions for the structure
functions at small $x\ll x_s\sim T/Q$ :
  \beq F_1&=&\frac{3N^2T^2}{16\Gamma^2(1/3)}\left(\frac{k}{3}
   \right)^{{2}/{3}}, \label{83} \\[0.2cm]
   F_L&\equiv &F_2-2xF_1=\frac{N^2Q^2x}{96\pi^2}\,,
   \label{84} \eeq
which represent our main result in this paper. Although strictly valid
only for $x\ll x_s$, these results remain parametrically correct also in
the transition region at $x\simeq x_s$. For $x\gg x_s$, on the other
hand, the structure functions are negligibly small, as discussed in Sect.
3.

To render the above results more transparent, it is convenient to rewrite
them in terms of the conventional variables for DIS, $x$ and $Q^2$, and
to also introduce the transverse structure function $F_T\equiv 2xF_1$,
such that $F_2=F_T+F_L$. Then Eqs.~(\ref{83}) and (\ref{84}) imply the
following parametric estimates:
  \beq\label{FTL} F_T(x,Q^2)&\sim& N^2\,\frac{T^2}{x}\,\left(\frac{x^2Q^2}{T^2}
  \right)^{2/3}\,,\nn[0.2cm]
  F_L(x,Q^2)&\sim& N^2\,\frac{T^2}{x}\,\left(\frac{x^2Q^2}{T^2}
  \right)\,,\eeq
which show that, in the very small--$x$ regime at $x \ll T/Q$, the
longitudinal structure function is negligible as compared to the
transverse one, $F_L\ll F_T$, and thus, somehow surprisingly, an analog
of the Callan--Gross relation applies: $F_2\simeq 2xF_1$. This looks
surprising since it is quite different from what happens in the case
where the target is a single hadron, at either weak coupling\footnote{In
QCD at weak coupling, the Callan--Gross relation holds only in the
Bjorken scaling regime at relatively large $x$, where the structure
functions are dominated by the valence quarks and depend very weakly upon
$Q^2$.} \cite{SATreviews}, or strong coupling \cite{dis,Hatta:2007he},
where in the high--energy limit $F_L$ and $F_T$ are parametrically of the
same order.

\section{Saturation and the partonic structure of the plasma}
\setcounter{equation}{0}

The results of the last two sections are conveniently described using
Fig. 3 where $\tau\equiv \ln 1/x$ is the `rapidity' and $\rho\equiv\ln
(Q^2/T^2)$. For a given $\rho\gg 1$ and values of $\tau$ below the {\em
saturation line} $\tau_s(\rho)=\rho/2$, meaning $x\gg x_s\simeq T/Q$, the
structure functions are {\em extremely} small (cf. Eq.~(\ref{Ftunnel}) in
Appendix B),
 \beq\label{Fsmalltau}
F_{L,\,T}\,\sim\,{N^2Q^2x}\, \exp\big\{-c(x/x_s)^{1/2}\big\}\,\propto
\,\exp\big\{-c\,\rme^{(\tau_s-\tau)/2}\big\}\qquad {\rm for}\qquad \tau <
\tau_s(\rho)\,,
 \eeq
while for values of $\tau$ significantly above that line ($x\ll x_s$) the
structure functions take on the values given in (\ref{83}) and
(\ref{84}). The transition between these two regimes when crossing the
saturation line is expected to occur within a rapidity interval $\Delta
\tau\sim \order{1}$.

\begin{figure}[t]
\begin{center}
\includegraphics[width=10.cm,height=8cm,bb=100 200 800 700]{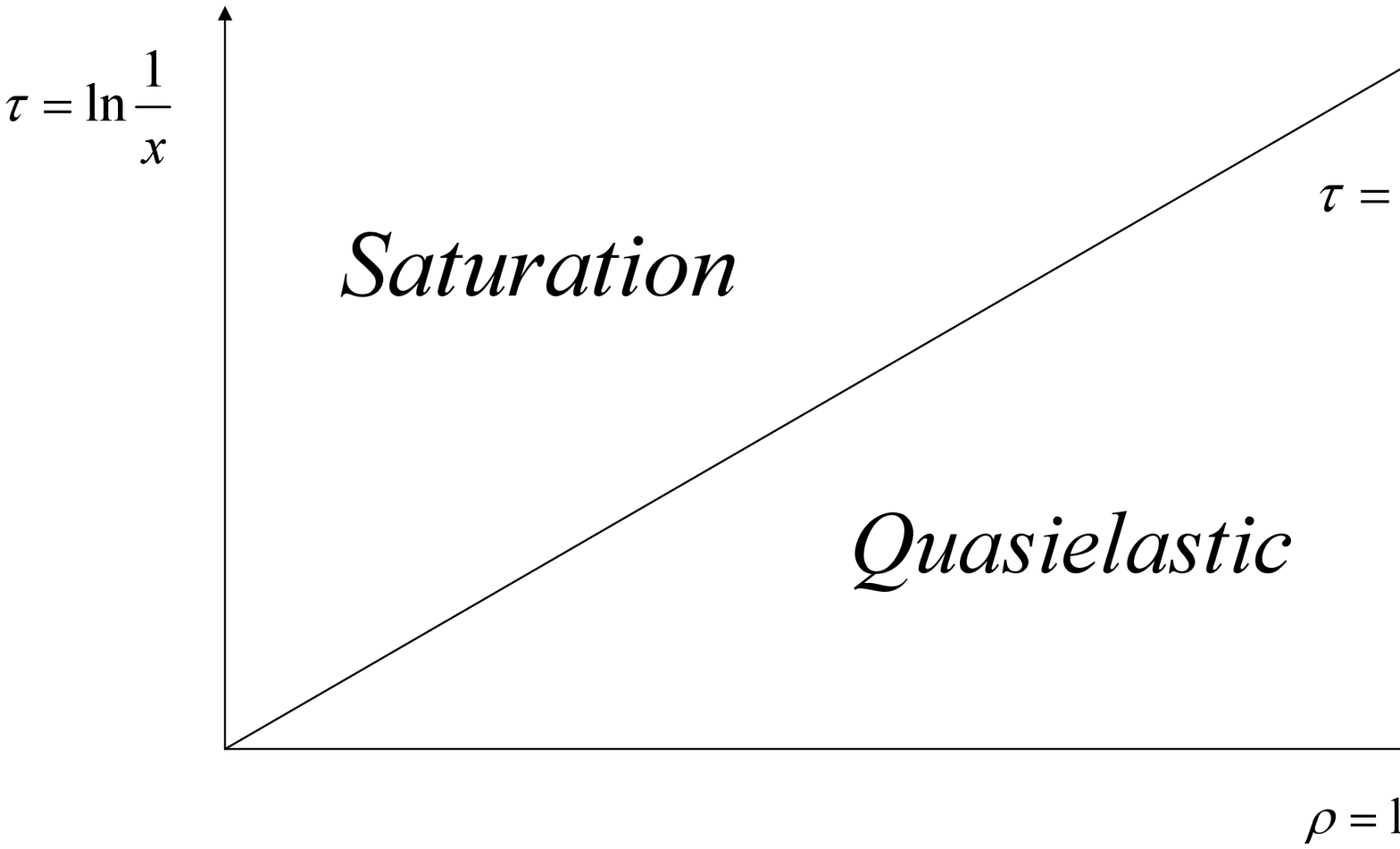}
\caption{\sl Proposed `phase diagram' for DIS off a ${\mathcal N}=4$
SYM plasma at high energy and
strong coupling.
 \label{phase2}}
\end{center}
\end{figure}

The saturation line can equivalently rewritten as $\rho_s(\tau)=2\tau$,
and then the estimates (\ref{Fsmalltau}) for the structure functions at
small $\tau$ are tantamount to
  \beq\label{Flargerho} F_i\,\sim\,
\exp\big\{-c(Q/Q_s)^{1/2}\big\}\,\sim
\,\exp\big\{-c\,\rme^{(\rho-\rho_s)/2}\big\}\qquad {\rm for}\qquad \rho
> \rho_s(\tau)\,,
 \eeq
with the {\em saturation momentum}
 \beq\label{Qsat}
  Q_s^2(\tau)\,\equiv\, T^2\,\rme^{\rho_s}\,=\,T^2\,\rme^{2\tau}
  \,=\,\frac{T^2}{x^2}\,.\eeq
Such a small value for $F_i$ for large $Q^2\gg Q^2_s(\tau)$ is
qualitatively consistent with previous calculations of the dilaton
structure functions at strong coupling \cite{dis,Hatta:2007he}, although
some quantitative differences remain. In these previous works, one has
found that the higher--twist terms dominate the dilaton structure
functions at large $Q^2$, thus yielding a fast decrease with $Q^2$, which
is however {\em power--like}, $F_i(x,Q^2)\propto (1/Q^2)^{\Delta}$ with
$\Delta \ge 1$, rather than exponential as predicted by
Eq.~(\ref{Flargerho}). Some of these higher--twist contributions are
naturally absent from the present analysis, since suppressed in the
large--$N$ limit. (This is the case for the diffractive processes
considered in Ref. \cite{Hatta:2007he}, which in the present framework
would correspond to multiple scattering off a {\em same} thermal
quasiparticle. The corresponding scattering amplitude starts at order
$1/N^4$, and hence it is suppressed at large $N$ even after
multiplication by the number $\sim N^2$ of thermal degrees of freedom.)
On the other hand, the higher--twist contributions due to protected
operators, as discussed for a dilaton target in Ref. \cite{dis}, would
survive in large--$N$ limit, but they are removed by the requirement of
energy--momentum conservation. (Being homogeneous in the four physical
dimensions, the plasma cannot transmit any energy or momentum via a
single scattering.) We interpret this smallness of $F_i$ at relatively
large $x$ to mean that for $x\gg x_s= T/Q$ there are hardly any
point--like excitations (partons) in the SYM plasma.

In what follows, we shall rather focus on the more interesting situation
at $x\lesssim x_s$, or large rapidity $\tau\gtrsim \tau_s(\rho)$, where
partons {\em do} exist, as we shall see. A natural place to look for a
partonic interpretation is at the level of the sum rules (\ref{sum1}) and
(\ref{sumL}). By inspection of our previous estimates (\ref{FTL}) for the
structure functions, it is easy to check that \texttt{(i)} the integrals
in Eqs.~(\ref{sum1}) and (\ref{sumL}) are dominated by values of $x$ of
order $x_s$ (for which the transverse and longitudinal structure
functions are of the same order of magnitude), and \texttt{(ii)} the
results of these integrations are of the right order of magnitude, namely
of $\order{N^2T^4}$, to ensure the fulfillment of the sum rules. For
instance, for  Eq.~(\ref{sum1}) we can write
 \beq \mathcal{E}\,= \,18T^2\int_0^1\rmd
 x\,F_2(x,Q^2)\sim T^2\,xF_2(x,Q^2)\Big|_{x=T/Q}\,, \label{87} \eeq
where $xF_2(x,Q^2)\sim N^2T^2$ when $x\simeq T/Q$, as manifest from
Eq.~(\ref{FTL}) (recall that $F_2=F_T+F_L$). Ours results in
Eqs.~(\ref{83})--(\ref{84}) are not accurate enough to also check the
numerical coefficients in front of the sum rules (this would require a
more precise study of the transition region at $x\sim x_s$). But the
parametric estimates in Eq.~(\ref{FTL}) are sufficient for our present
purpose, which is to develop a partonic picture for the strongly coupled
plasma.

Before we proceed, let us first recall the interpretation of the
structure function $F_2$ in the more familiar context of perturbative
QCD. In that case, $F_2(x,Q^2)$ is a dimensionless quantity interpreted
as the {\em quark distribution} in the proton target, i.e., the number of
quarks which are localized in impact parameter space within an area $\sim
1/Q^2$ fixed by the resolution of the virtual photon, and which are
distributed in longitudinal phase--space within a unit of rapidity
($\Delta\tau\sim 1$) around the rapidity $\tau=\ln(1/x)$ fixed by the
Bjorken variable. In what follows, we shall boldly propose a similar
interpretation for the strongly--coupled plasma, and then critically
examine the most sensible points in our proposal.

Unlike the proton, or dilaton, structure functions, which are
dimensionless, the plasma structure functions $F_i$ as computed in this
paper have dimensions of $({\rm area})^{-1}$. This makes it natural to
try and relate these functions to the density of partons per unit area in
the transverse plane $(x,y)$ (the impact parameter space). The ${\mathcal
R}$--current with $Q^2\gg T^2$ probes an area $\sim 1/Q^2$ much smaller
then the typical area $\sim 1/T^2$ covered by a `thermal quasiparticle'
in the plasma --- i.e., a typical thermal excitation with energy and
momentum of order $T$ (in the plasma rest frame). This means that the
current can see `inside' a quasiparticle, and thus probe its elementary
constituents, or `partons'. More precisely, it will simultaneously
scrutinize inside all the quasiparticles located within one {\em
coherence length} in the longitudinal direction $z$. The notion of
coherence length is particularly important for what follows, and so is
also the choice of an appropriate Lorentz frame in which the parton
interpretation makes sense. So, let us open a parenthesis at this point,
in order to better explain these concepts :

{\texttt{(i)}} The partonic picture makes sense in a frame where the
current has low energy and a relatively simple internal structure, so
that it can act as a probe of the target. Here, it will be convenient to
use the Breit frame where the ${\mathcal R}$--current is a standing wave.
Namely, if one boosts the plasma by an amount $\eta$ where $\cosh \eta\!=
\!q/Q$, then in this boosted frame the current has time and $z$--momentum
components $\omega^{\prime}\!=\!0$ and $q^{\prime}\!=\!Q$. This current
is naturally absorbed by partonic constituents of the boosted plasma
having momenta of order ${Q}$. Indeed, the partons participating in the
collision have a longitudinal momentum fraction $x$, and thus a typical
momentum $p'_z \!\sim \!x(T\cosh \eta)\!\sim\! Q$ in the boosted frame.

{\texttt{(ii)}} Before boosting, the current correlator (\ref{1}) is
sensitive to longitudinal distances $\Delta z\lesssim 2q/Q^2$, as is
suggested by writing the space--time dependence in the integral there as
 \beq\label{exp}
 \rme^{-i\omega t+iqz}\,\simeq\,\rme^{-iq(t-z)+iQ^2t/2q}\,, \eeq
where we have used $Q^2\simeq 2q(q-\omega)$ for $q^2\gg Q^2$. This shows
that the integration in (\ref{1}) can extend over the coherence time
$\Delta t_c\sim 2q/Q^2$, corresponding to a coherence length $\Delta
z_c\sim 2q/Q^2$ in the plasma rest frame. After the boost, this length
gets Lorentz contracted (note that the current is decelerated) down to a
value $\Delta z'_c\sim (2q/Q^2)/\cosh \eta\sim 1/Q$.

Let us now return to Eq.~(\ref{87}) and try to interpret this sum rule in
the Breit frame. In the l.h.s., we would like to construct the energy
density per unit area, $\rmd E'/\rmd^2b$, of the region of the plasma
which is explored by the ${\mathcal R}$--current in this boosted frame.
As a component of the second--rank tensor $\Theta_{\mu\nu}$, the
(three--dimensional) energy density $\mathcal{E}$ transforms in the boost
by a factor $(\cosh \eta)^2$. The ${\mathcal R}$--current probes a slice
of the plasma with longitudinal extent $\Delta z'_c$ in the boosted
frame. Hence, $\rmd E'/\rmd^2b\sim (\mathcal{E} \cosh^2 \eta)\Delta z'_c
\simeq \mathcal{E} (2q/Q^2) \cosh \eta$. Multiplying both sides of
(\ref{87}) by $(2q/Q^2) \cosh \eta$, one gets
 \beq  \label{89} \frac{\rmd E'}{\rmd^2b}
 \sim xT\cosh \eta \left(\frac{1}{x}F_2(x,Q^2)\right)_{x=T/Q}\,.
 \eeq
As before mentioned, the quantity $xT\cosh \eta\sim Q$ in the r.h.s. is
the longitudinal momentum of the constituent (parton) which interacts
with the ${\mathcal R}$--current. It is therefore natural to interpret
 \beq \frac{1}{x}F_2(x,Q^2)\Big|_{x=T/Q} \sim
 \frac{\rmd n}{\rmd^2b}\Big|_{x=T/Q}\,,  \label{90} \eeq
as the number of partons per unit area within a longitudinal slice of the
plasma, with the width of the slice equal to the coherence length (which
is $1/Q$ in the Breit frame, and $q/Q^2$ in the plasma rest frame). This
interpretation, which here has been inferred from the sum rule
(\ref{87}), is in fact natural in view of the standard partonic
interpretation of $F_2$ at weak coupling, as alluded to before. The only
new feature with respect to the case where the target is a single
proton\footnote{There is no such a factor in the case of a single--hadron
target since there the whole longitudinal extent of the hadron lies
within one coherence length for the virtual photon.} is the factor $1/x$
in the l.h.s.: this is a Lorentz--invariant measure of the amount of
matter in the plasma in the longitudinal slice explored by the current.
Namely, this has been generated as (say, in the plasma rest frame):  $T
\Delta z_c\sim 1/x$, where $T$ is the density of quasi--particles per
unit length and $\Delta z_c$ is the longitudinal extent of the
interaction region.

For what follows, it is useful to notice that the parton density in the
r.h.s. can be equivalently written
   \beq\label{long}
 \frac{\rmd n}{\rmd^2b}\,= \,\Delta z'\, \frac{\rmd n}{\rmd z'
 \rmd^2b}\,\simeq\, \, \frac{\rmd n}{p^{\prime}_z \rmd z'
 \rmd^2b}\,=\,\frac{\rmd n}{\rmd\tau\rmd^2b}\,,\eeq
where $\Delta z'\sim 1/Q$ is the longitudinal extent of the slice in the
boosted frame and $p'_z\sim Q$ (the $z$--momentum of a struck parton) is
the same as $1/\Delta z'$, as it should by the uncertainty principle. In
writing the last equality, we have identified the rapidity interval
$\rmd\tau=p^{\prime}_z \rmd z'$.

By using Eq.~(\ref{90}) together with the previous estimate (\ref{FTL})
for $F_2=F_T+F_L$, one finds
  \beq
 \frac{\rmd n}{\rmd\tau\rmd^2b}\,\sim \,N^2Q^2\qquad\mbox{for}\qquad
 x\sim T/Q\,, \label{nsat} \eeq
which, remarkably, has the same parametric form as in a weakly--coupled
gauge theory \cite{SATreviews}, and hence it admits a similar physical
interpretation. Namely, when interpreted in the Breit frame,
Eq.~(\ref{nsat}) gives the total number of partons (per unit area) having
a longitudinal momentum fraction equal to $x$ (with $x\lesssim T/Q$) and
with transverse momenta $p_\perp \lesssim Q$. Since this number appears
to be of order $N^2Q^2$, we conclude that there is a number of order one
of partons (of a given color) per unit of phase--space:
 \beq
 \frac{1}{N^2}\, \frac{\rmd n}{\rmd\tau\rmd^2 p_\perp \rmd^2 b_\perp}
 \,\simeq\,1\qquad\mbox{for}\qquad
 p_\perp\lesssim Q_s(x)=T/x\,. \label{phisat} \eeq
(The factor $\rmd\tau$ plays the role of $p^{\prime}_z \rmd z'$, cf.
Eq.~(\ref{long}), so the above phase--space density is an {\em occupation
number}, in the proper, three--dimensional, sense.) This is similar to
pQCD in the sense that the parton occupation number saturates at
sufficiently low transverse momenta, below a critical scale $Q_s(x)$
which grows like a power of $1/x$. In QCD, saturation is a reflection of
unitarity in a corresponding scattering process. Where is the unitarity
limit here? Viewed on the gravity side of the AdS/CFT correspondence the
gravitational wave $A_\mu$ induced by the ${\mathcal R}$--current is
completely absorbed at the horizon of the black hole and that absorption
takes place over a time less than or equal to the coherence time, $1/xT$,
of the wave. This is, in effect, a unitarity limit for scattering of the
gravitational wave (very much similar to the corresponding limit for
dipole scattering in the familiar dipole factorization for DIS at high
energy \cite{SATreviews}).

Recently, the phenomenon of parton saturation in relation with the
unitarity limit for DIS has also been identified at {\em strong
coupling}, in the case where the target is a single `hadron' (a dilaton)
\cite{Hatta:2007he}. Interestingly, our above result (\ref{Qsat}) for the
saturation momentum of the plasma is consistent with the corresponding
result for a single hadron in Ref. \cite{Hatta:2007he}, once the assumed
structure of the SYM plasma in terms of quasiparticles is taken into
account. Namely, the quantity $Q_s^2$ is proportional to the density of
partons per unit area in impact parameter space. In the case of a single
dilaton, Ref. \cite{Hatta:2007he} has found
 \beq
  Q_s^2(x)\,=\, \frac{\Lambda^2}{xN^2} \qquad
  \mbox{(one dilaton target)}\,, \label{Qshigh}
  \eeq
with $1/\Lambda$ a measure of the dilaton transverse size. When moving to
the plasma, the dilaton gets replaced by thermal quasiparticles with
individual size $\sim 1/T$. To account for the degrees of freedom
relevant to DIS, one must sum over color (this yields a factor $N^2$) and
also over the number of quasiparticles within a longitudinal slice of
width $\Delta z_c \sim q/Q^2$ in the plasma rest frame (which gives an
additional factor $T \Delta z_c\sim 1/x$). After replacing $\Lambda\to T$
in Eq.~(\ref{Qshigh}) and putting these various factors together, we end
up with the previous result, Eq.~(\ref{Qsat}), as anticipated. This is an
important check of the internal consistency of our proposed partonic
description --- it comforts our idea that DIS off the plasma at high
$Q^2\gg T^2$ should measure the internal constituents of the thermal
quasiparticles composing the plasma. This check is particularly
non--trivial in view of the fact that the unitarization mechanisms at
work appear to be very different in the two cases --- disappearance of
the potential barrier for the plasma case, respectively, diffractive
scattering via multiple graviton exchanges in the dilaton case.

We now turn to the case $x\ll T/Q$, which turns out to be quite subtle.
Previously, we argued that the typical interaction time is of the order
of the coherence time $\Delta t_c\sim q/Q^2$ of the incoming current.
This argument, however, ceases to be valid at very high energy, where the
gravitational wave gets absorbed (reaches the horizon) on a time scale
shorter than $\Delta t_c$. A heuristic way to understand this is to
recall that, when the energy is so high that the barrier has disappeared,
cf. Fig. 1c (namely, for $qT^2/Q^3\gg 1$), the $F_1$ structure function
becomes independent of $Q^2$, as manifest on Eq.~(\ref{83}). On the other
hand, the definition (\ref{1}) of the current--current correlator
involves an explicit dependence upon $Q^2$, via the exponential factor
inside the integrand, conveniently written as in Eq.~(\ref{exp}). The
only way for this dependence to disappear at high energy is that the
integral over $t$ in Eq.~(\ref{1}) be cut at some time which is
considerably shorter than the coherence time $\Delta t_c$. This requires
the lifetime of the gravitational wave (before being absorbed by the
black hole) to be shorter than $\Delta t_c$. For a given energy $q$, this
lifetime can be estimated as $q/Q_s^2(q)$, since $Q^2=Q_s^2(q)$ is the
smallest value of $Q$ for which Eqs.~(\ref{key}) and (\ref{ai}) still
have a $Q$--dependence. Here, $Q_s^2(q)$ is the saturation momentum
expressed as a function of $q$, and is obtained from the condition
$qT/Q_s^2=Q_s/T$ (the condition to lie on the saturation line in Fig. 3)
as $Q_s^2(q)=(qT^2)^{2/3}$. Note that $q/Q_s^2(q)$ is indeed much smaller
than $q/Q^2$ in this high energy regime.

Because of this short lifetime of the high--energy gravitational wave, we
believe that the quantity $F_1(x,Q^2)\simeq(1/2x)F_2(x,Q^2)$ (which, we
recall, is the dominant structure function when $x\ll T/Q$) is actually
being determined by interactions of the ${\mathcal R}$--current with
partons of size $1/Q_s^2(q)$, rather than with partons of size $1/Q^2$.
Indeed, Eq.~(\ref{83}) implies
 \beq F_1(x,Q^2)\,=\,F_1(x_s,Q^2_s) \qquad\mbox{where}\quad
 Q_s^2(q)=(qT^2)^{2/3} \quad\mbox{and}\quad x_s(q)=T/Q_s(q)\,.\eeq
Let us give another argument leading to the same conclusion. We recall
that in DIS in the QCD dipole picture, and in the rest frame of the
target, the size of the dipole emerging from the electromagnetic current
expands with time as $\Delta x_\perp\sim \sqrt{t/2q}$
\cite{Farrar:1988me}, so that it takes a time $t\sim 2q/Q^2$ for this
dipole to reach a size $\Delta x_\perp\sim 1/Q$. Assume that a similar
estimate applies for the SYM system emerging from the ${\mathcal
R}$--current (the analog of the QCD color dipole); then, after a time
$t_s\sim 2q/Q_s^2(q)$, which is the lifetime of this SYM system before
being absorbed by the plasma, its size gets only as big as $\Delta
x_\perp\sim 1/Q_s(q)$, which indicates once again that the partons at
scale $Q_s(q)$ are the relevant degrees of freedom.

We shall conclude this discussion, and also the paper, with a critical
analysis of the main assumptions that we have made in reaching
Eqs.~(\ref{90}) and (\ref{nsat}) --- the equations at the basis of our
partonic interpretation. \texttt{(i)} We have assumed the plasma to be
made of constituents (`quasiparticles') having momenta on the order of
$T$ (in the plasma rest system) when measured on a resolution scale $T$.
The fact that entropy density and energy density scale as $N^2T^3$ and
$N^2T^4$ suggest that this is the case, but this understanding is,
perhaps, not completely clear. Note that, for the present purposes, we
did not need to specify the actual nature of these `quasiparticles',
which at strong coupling would be a most difficult task. \texttt{(ii)} We
have also assumed that the ${\mathcal R}$--current directly measures
individual constituents at scale $Q$. In QCD the electromagnetic current
provides such a measurement in the leading order renormalization group
formalism. At next--to--leading order, ambiguities occur in separating
the measured partons from the probe; however, these ambiguities are
effects of order $\alpha(Q^2)$ and cannot affect general conclusions as
to numbers of partons in a hadron or plasma. In SYM, we have taken the
coupling large so that the separation between the probe and the partons
to be measured is not sharp anymore. In reaching (\ref{90}) and
(\ref{nsat}) we have assumed that, up to factors of order one, the
${\mathcal R}$--current couples to individual constituents of the plasma
and that this coupling is not strongly renormalized. Because of these
subtleties we feel that our results have to be taken with caution, and
that a deeper understanding of the partonic structure of the plasma in
strong coupling SYM is highly desirable.

\section*{Acknowledgments}
 We would like to thank Iosif Bena for useful discussions. The work of
A.H.~M. is supported in part by the US Department of Energy. The work of
E.~I. in supported in part by Agence Nationale de la Recherche via the
programme ANR-06-BLAN-0285-01.

 \appendix
\section{Structure functions: Definitions and sum rules}
\setcounter{equation}{0}

In this appendix we remind the reader of the tensor structure of
$R_{\mu\nu}$ and derive a sum rule relating the expectation of energy
momentum tensors in the plasma to deep inelastic scattering on the
plasma. $R_{\mu\nu}$ is defined in Eq.~(\ref{1}), The tensor structure
must be given in terms of $q^\mu$ and the plasma four-velocity
$n^\mu=(n_t,n_x,n_y,n_z)$, with $n^\mu=(1,0,0,0)$ corresponding to the
plasma at rest. Then current conservation plus the symmetry property
$R_{\mu\nu}(q)=R_{\nu\mu}(-q)$ imply the following general structure
(with $\eta_{\mu\nu}=(-1,1,1,1)$) :
  \beq R_{\mu\nu}=
  \left(\eta_{\mu\nu}-\frac{q_\mu q_\nu}{Q^2} \right)R_1+
  \left[n_\mu n_\nu-\frac{n\cdot q}{Q^2}
 (n_\mu q_\nu + n_\nu q_\mu)+
 \frac{q_\mu q_\nu}{(Q^2)^2}(n\cdot q)^2 \right] R_2\,. \label{A.1} \eeq
The two scalar functions $R_1$ and $R_2$ depend upon the two invariants
$Q^2$ and $x$ introduced in Eq.~(\ref{Qxdef}), and they are even
functions of $x$. We define the DIS structure functions as (note that
$n\cdot q=-\omega$ is negative)
   \beq F_1&\,=\,&\frac{1}{2\pi}\ {\rm Im} R_1, \label{F1} \\[0.2cm]
  F_2&\,=\,&\frac{-(n\cdot q)}{2\pi T}\ {\rm Im} R_2 \label{F2}\,. \eeq

Writing the energy--momentum tensor of the plasma as
 \beq\label{theta}
\Theta_{\mu\nu}=(\eta_{\mu\nu}+4n_\mu n_\nu)\,\frac{\mathcal{E}}{3}
\qquad \mbox{with} \qquad
 \mathcal{E}\,=\,\frac{3N^2\pi^2T^4}{8}\,, \eeq
we can rewrite the leading--twist results in Eq.~(\ref{R12LT}) as
 \beq\label{R121}
 R_1^{(1)}\,=\,\frac{\mathcal{E}}{15T^2x^2}\,,\qquad R_2^{(1)}\,=\,
 \frac{4\mathcal{E}}{9Q^2}\,. \eeq
This rewriting makes it clear that the calculation of the leading--twist
contribution to $R_{\mu\nu}$ in Sect. 3 amounts to computing the
coefficients of the energy--momentum tensor in the operator product
expansion for the current--current correlator.

To deduce the sum rules (\ref{sum1}) and (\ref{sumL}), we shall write
$z=1/x$ and assume the standard analytic structure for the functions
$R_i(z)$ in the complex $z$ plane. Namely, $R_i(z)$ is an analytic
function everywhere in the complex plane except for two cuts along the
real axis (from $z=-\infty$ to $z=-1$ and, respectively, from $z=1$ to
$z=\infty$). Then our previous results in Eq.~(\ref{R121}) express the
dominant behaviour of $R_i(z)$ near $z=0$. Using this information
together with Eq.~(\ref{R121}), we can successively write
\beq
\frac{4\mathcal{E}}{9Q^2}&\,=\,&\oint \frac{\rmd z}{2\pi i} \frac{R_2}{z}
=2\int_1^\infty\frac{\rmd z}{2\pi i}\frac{2i \,{\rm Im}R_2}{z}\nn[0.2cm]
 &\,=\,& \frac{2}{\pi}\int_0^1\rmd x \ \frac{{\rm
Im}R_2}{x} \,=\,\frac{8T^2}{Q^2}\int_0^1 \rmd x F_2(x,Q^2)
 \eeq
where the contour in the first integral is a small circle surrounding the
origin. This is then distorted in the complex plane in such a to wrap
around the two branch cuts which give equal contributions. (We assume the
integrand to vanish sufficiently fast as $|z|\to\infty$ to be able to
neglect the contributions of the large circles closing the contour.)
Similarly,
\beq \frac{\mathcal{E}}{15T^2}&=&\oint \frac{\rmd z}{2\pi i}
\frac{R_1}{z^3} =2\int_1^\infty\frac{\rmd z}{2\pi i}\frac{2i\, {\rm
Im}R_1}{z^3}\nn[0.2cm] &=&\frac{2}{\pi}\int_0^1\rmd x \,x\ {\rm Im}R_1
=4\int_0^1 \rmd x\,x F_1(x,Q^2)
\eeq

\section{Low energy: tunnel effect}
\setcounter{equation}{0}

In this Appendix we shall use WKB techniques to estimate the probability
for inelastic scattering via tunnel effect in the intermediate energy
regime at $Q\ll q\ll Q^3/T^2$, where the potential barrier is high. The
argument turns out to be non--trivial because the imaginary part of the
classical solution --- which, we recall, is the measure of inelasticity
in the scattering --- gets built via a `double--tunnel effect' (see
below), for which the WKB approximation is generally not reliable. Yet,
as we shall later argue, in the present setup this approximation should
be reliable for the imaginary part of the solution.

We shall focus on the longitudinal wave (the corresponding discussion of
the transverse wave is entirely similar) and use the wave equation in
Schr\"odinger form, cf. Eqs.~(\ref{ad})--(\ref{pot}). We shall construct
our global approximation for $\psi$ by matching approximate solutions
valid in three different domains: \texttt{(i)} $u$ close to zero,
\texttt{(ii)} $u$ inside the potential barrier in Fig. 1a, and
\texttt{(iii)} relatively large $u$, on the right side of the barrier. As
usual, the imaginary part in the solution will be generated by the
condition that $\psi(u)$ be a purely outgoing wave at large $u\sim
\order{1}$.

\texttt{(i)} For relatively small $u$, the potential in Eq.~(\ref{pot})
can be approximated as
 \beq
 V\simeq\frac{1}{u}\left[-\frac{1}{4u}+K^2\right]\qquad\mbox{for}\qquad
 0\,\le \,u\,\ll\,K/k\,.
   \label{VI} \eeq
Then the general solution can be written as:
  \beq \psi(u)
  =C_1\sqrt{u}\,\mathrm{K}_0(2K\sqrt{u})\,+\,
  C_2\sqrt{u}\,\mathrm{I}_0(2K\sqrt{u})\,,
  \label{psiI} \eeq
where the coefficient $C_1$ is fixed by the boundary condition at $u=0$,
Eq.~(\ref{bc}), as $C_1=-2k^2\mathcal{A}_L(0)$. This approximation is
similar to the zeroth order perturbative solution (\ref{a0sol}) in Sect.
3 except that, now, the coefficient $C_2$ in front of $\mathrm{I}_0$ is
allowed to be non--zero because of the different behaviour assumed at
large $u$. The imaginary part, $\rm{Im}\,C_2$, of this coefficient is the
quantity that we are primarily interested in, because this quantity
determines, via Eq.~(\ref{actioncl}), the imaginary part of the
`on--shell' action.

\texttt{(ii)} For values of $u$ inside the potential barrier, $u_1 < u <
u_2$ with $u_1\simeq 1/(4K^2)$ and $u_2\simeq K/k$ the two classical
turning points in Fig. 1a, the solution can be constructed via the WKB
approximation, which yields
  \beq \label{psiII} \psi(u)\simeq
  \frac{1}{\sqrt{V(u)}}\left[
  C_3\exp\left\{-\int_{u_1}^{u}\rmd u' \sqrt{V(u')}\right\}
  +
  C_4\exp\left\{\int_{u_1}^{u}\rmd u'
  \sqrt{V(u')}\right\}\right]\,.\eeq
In applications of the WKB technique to the tunnel effect, the analog of
the second term in the equation above is generally omitted, since beyond
the accuracy of this approximation. However, in so far as the imaginary
part of the solution is concerned --- which, we recall, is our main
interest here --- the inclusion of this term is both essential and
justified, as we shall later argue.

The approximate solutions (\ref{psiI}) and (\ref{psiII}) have a common
validity range at $u_1 < u\ll K/k$, and thus can be matched with each
other in this window. By also using the asymptotic behaviour of the
modified Bessel functions, as valid for $u\gg u_1$, one finds
 \beq\label{C1234}
 C_1=\frac{2}{\sqrt{\pi}}\, C_3\,,\qquad
 C_2={2}{\sqrt{\pi}}\ C_4\,.\eeq

\texttt{(iii)} For $u\gg u_2$, the WKB solution is similar to the one
constructed in Sect. 4 and reads
  \beq \label{psiIII} \psi(u)\simeq
  \frac{C}{\sqrt{-V(u)}}\,
  \exp\left\{i\int_{u_2}^{u}\rmd u' \sqrt{-V(u')}\right\}\,,
  \eeq
where we have selected only the outgoing wave, i.e., the one propagating
towards the black hole. (This corresponds to choosing $c_5=0$ in
Eq.~(\ref{66}).) The above coefficient $C$ is the same as $c_4$ in
Eq.~(\ref{66}), but its precise value is irrelevant here (it would merely
determine the normalization of the wave near $u=1$, cf. Eqs.~(\ref{60})
and (\ref{c34})). Rather, what matters is the relative normalization of
the coefficients $C_3$ and $C_4$ in the solution (\ref{psiII}) inside the
barrier, which in turn is fixed by matching Eqs.~(\ref{psiII}) and
(\ref{psiIII}) near $u=u_2$. This matching cannot be done by directly
comparing these two solutions, as they have no overlap with each other.
Yet, the proper matching procedure is standard in the WKB literature
\cite{landau,Bender} (this requires a study of the exact behaviour near
$u_2$, which can be done by linearizing the potential and then
recognizing the Airy equation), and here we shall simply list the result:
 \beq\label{C34D}
 C_3=\frac{C}{\sqrt{D}}\ \rme^{-i\pi/4}\,,\qquad
 C_4=\frac{i}{2}\,\sqrt{D}\,C\,\rme^{-i\pi/4}\,=\,\frac{i}{2}\,DC_3,\eeq
where $D$ is the WKB attenuation factor (in the usual context of quantum
mechanics, this describes the decrease in the intensity $|\Psi|^2$ of the
wavefunction after passing the potential) :
 \beq\label{DWKB}
 D\,\equiv\,\exp\left\{-2\int_{u_1}^{u_2}\rmd u \sqrt{V(u)}\right\}\,.
 \eeq
By comparing Eqs.~(\ref{C1234}) and (\ref{C34D}) one finds
 \beq\label{C12}
 C_2\,=\,i\,\frac{\pi}{2}\,D\, C_1\,,\eeq
which implies the following behaviour for (\ref{psiI}) near the boundary
at $u=0$ (recall that $\psi(u)\simeq\sqrt{u}\,a(u)$ for small $u$) :
  \beq a(u)\,\simeq\,
  k^2\mathcal{A}_L(0)\,
  \big[\ln K^2+(\ln u +2\gamma) \,-\,i\pi D\,\big]\,.
  \label{atunnel} \eeq
This is our main result here. It shows that, in the presence of a high
potential barrier, the imaginary part of the solution near $u=0$ gets
built via a {\em double--tunnel effect}. This is `double' since the
relative strength of the imaginary part versus the real part is $D$, and
not $\sqrt{D}$. This result is in fact natural: this imaginary part is
the feedback of the absorption taking place near $u=1$ on the
gravitational perturbation at the Minkowski boundary. First, the incoming
perturbation, which is purely real, has to cross the barrier to approach
the black hole, then, after the scattering takes place near $u=1$, the
imaginary part thus generated in the solution must propagate backwards
and cross the barrier once again, before being measured (in the form of
DIS structure functions) at $u=0$.

By using Eq.~(\ref{atunnel}) together with the corresponding equation for
the transverse sector, one can finally compute the DIS structure
functions generated through tunneling in this low--energy (or
low--temperature) regime. One thus finds quite similar expressions for
the longitudinal ($F_L=F_2-2xF_1$) and transverse ($F_T=2xF_1$) structure
functions :
 \beq
    F_i\,\simeq\,\frac{N^2Q^2x}{32\pi^2}\,D_i\qquad(i\,=L,\,T)\,.
   \label{Ftunnel} \eeq
It is easy to check that the integral in Eq.~(\ref{DWKB}) is dominated by
the region in $u$ where the potential can be simplified to $V(u)\simeq
(K^2-k^2u^2)/u$, which is the same as the potential (\ref{tpot}) in the
transverse sector and for $u\ll 1$. Then, the attenuation factor is
essentially the same (to leading exponential accuracy) for both the
longitudinal and the transverse waves, and can be estimated as
 \beq D\sim \exp\big\{-c(K^3/k)^{1/2}\big\}\qquad\mbox{with}\qquad
 c=\frac{2\Gamma^2(1/4)}{3\sqrt{2\pi}}\,.\eeq
This explains the estimates
(\ref{Fsmalltau})--(\ref{Flargerho}) for $F_T$ and $F_L$.

Let us finally explain why, in the present context, we think that it was
justified to keep the second term in the WKB solution (\ref{psiII}).
Generally, this term is discarded in applications of the WKB method
\cite{landau,Bender} since it is exponentially suppressed as compared to
the first term there (recall that $C_4\propto DC_3$, cf. (\ref{C34D})),
and hence it is much smaller than the corrections to the prefactor in
that first term, which are only power--suppressed (in this case, by
rational powers of $k/K^3$). However, in the present problem, the first
exponential in (\ref{psiII}) is matched onto the {\em real} part,
$\propto C_1\,\mathrm{K}_0$, of the solution at small $u$; hence this
large exponential term is strictly real, and so would be all the
higher--order terms, neglected by the WKB approximation, which would
correct its prefactor. Accordingly, the second exponential in
(\ref{psiII}) is the only one which can develop an imaginary part, and
this imaginary part is therefore correct to WKB accuracy. To conclude,
the WKB approximation cannot be trusted for the real part of the
coefficient $C_4$ in (\ref{psiII}), but only for its imaginary part,
which is the quantity of interest for us here.


\section{The operator product expansion at weak coupling}
\setcounter{equation}{0}

In this Appendix we shall show that, when computed to lowest--order in
perturbation theory, the coefficient of the energy--momentum tensor in
the operator product expansion (OPE) of the current--current correlator
(\ref{1}) is exactly the same as the corresponding coefficient in the
strong--coupling limit, as implicitly computed in (\ref{R12LT}) (or in
Eq.~(\ref{R121})). This nonrenormalization property reflects the high
degree of supersymmetry of ${\mathcal N}=4$ SYM  (see, e.g.,
\cite{Arutyunov:2000ku}). In the perturbative calculation of the OPE to
follow, we shall keep only the operators which mix with the
energy--momentum tensor $T_{\mu\nu}$.


In ${\mathcal N}=4$ SYM, there are six scalars $\phi_m$ in the vector
representation of $SO(6)$ and four Weyl fermions $\psi_i$ in the
fundamental representation of $SU(4)$. The ${\mathcal R}$ symmetry
current corresponding to the generator $t^3={\rm diag}(1/2,-1/2,0,0)$ is
  \beq
 J^\mu=
 \frac{1}{2}(\psi^1\bar{\sigma}^\mu
 \psi^1-\bar{\psi^2}\bar{\sigma}^\mu
 \psi^2)+ \frac{1}{2}\left(\phi_6D^\mu \phi_5-\phi_5 D^\mu
 \phi_6+\phi_4D^\mu \phi_3-\phi_3D^\mu \phi_4 \right)\,, \eeq
where $\bar\sigma^{\mu}=(1,-\vec{\sigma})$, $\vec{\sigma}$ being the
Pauli matrices. By contracting fields with free propagators, it is
straightforward to derive the relevant part of the
 OPE:  \begin{align}
&i\int \rmd^4x \,\rme^{-iqx}
 J^\mu(x)J^\nu(0) \nonumber \\
 &=\frac{1}{Q^2}\left(\eta^{\mu\alpha}\eta^{\nu \beta}-\frac{q^\mu
 q^\alpha}{Q^2}\eta^{\nu\beta}-\eta^{\mu\alpha}\frac{q^\nu q^\beta}{Q^2}
 +\eta^{\mu\nu}
 \frac{q^\alpha
 q^\beta}{(Q^2)^2}\right)\left(\sum_{i=1,2} T_{\alpha\beta}^{\psi,i}+
 \sum_{m=3,4,5,6} T_{\alpha\beta}^{\phi,m}\right)\nonumber
 \\ &-\frac{1}{(Q^2)^2}\left(\eta^{\mu\nu}-\frac{q^\mu
 q^\nu}{Q^2}\right)q^\alpha q^\beta \sum_{m=3,4,5,6}T_{\alpha\beta}^{\phi,m}+
 \cdots\nonumber \\
 &=\frac{1}{Q^2}\left(\eta^{\mu\alpha}\eta^{\nu \beta}-\frac{q^\mu
 q^\alpha}{Q^2}\eta^{\nu\beta}-\eta^{\mu\alpha}\frac{q^\nu q^\beta}{Q^2}
 +\eta^{\mu\nu}
 \frac{q^\alpha
 q^\beta}{(Q^2)^2}\right) \left(\frac{1}{2}T_{\alpha\beta}^{\psi}+
 \frac{2}{3}T_{\alpha\beta}^{\phi}
 +({\rm nonsinglet\  terms})\right)\nonumber
 \\& -\frac{1}{(Q^2)^2}\left(\eta^{\mu\nu}-\frac{q^\mu
 q^\nu}{Q^2}\right)q^\alpha q^\beta \left(\frac{2}{3}
 T_{\alpha\beta}^{\phi}+({\rm nonsinglet\  terms)}\right)+\cdots\,.
 \label{sum}
 \end{align}
where in writing the second equality we have projected onto the $SU(4)$
singlet operators and denoted
 \beq T_{\alpha\beta}^{\phi}\equiv \sum_{m=1}^6
T_{\alpha\beta}^{\phi,m}\equiv \sum_{m=1}^6\phi_m iD_\alpha iD_\beta
\phi_m \eeq and \beq T_{\alpha\beta}^{\psi}\equiv \sum_{i=1}^4
T_{\alpha\beta}^{\psi,i} \equiv \sum_{i=1}^4
\frac{i}{2}\bar{\psi}_i(\bar{\sigma}_\alpha D_\beta+\bar{\sigma}_\beta
 D_\alpha)\psi_i\,. \eeq
These operators represent the energy--momentum tensors for scalar and
fermion fields, respectively. Under renormalization, they mix with the
total energy--momentum tensor, which also includes the respective
operator for the gluon fields and reads
\beq T_{\mu\nu} &=& F^a_{\mu\lambda}F^{a\lambda}_\nu +
 \frac{i}{2}\sum_{i=1}^4
 \bar{\psi}_i(\bar{\sigma}_\mu D_\nu+\bar{\sigma}_\nu
D_\mu)\psi_i+ \sum_{m=1}^6D_\mu \phi_mD_\nu \phi_m +\cdots \nonumber \\
&\equiv& T_{\mu\nu}^g+T_{\mu\nu}^\psi+T_{\mu\nu}^\phi\,. \eeq Their
mixing is governed by the anomalous dimension matrix for twist--two
operators. The eigenoperators of the anomalous dimension matrix are
\cite{Anselmi:1998ms} \beq T_{I}\equiv T^g+T^\psi+T^\phi, \qquad
 T_{II}\equiv -2T^g+T^\psi+2T^\phi, \qquad
 T_{III}\equiv -T^g+ 4T^\psi -6T^\phi\,. \eeq
The last two operators (unlike the former) have nonzero anomalous
dimensions. After decomposing the operators which appear in the OPE
(\ref{sum}) in terms of the above eigenvectors, i.e.,
 \beq
 \frac{1}{2}T^{\psi}+
 \frac{2}{3}T^{\phi}& =&
 \frac{1}{3}T_{I} +\frac{1}{6}T_{II},
 \nonumber \\
  \frac{2}{3}
 T^{\phi}&=&\frac{2}{15}T_{I}+
 \frac{2}{21}T_{II}-\frac{2}{35}T_{III}\,,
 \eeq
we finally get
 \beq\label{JJ} i\int \rmd^4x \rme^{-iqx}
 J^\mu(x)J^\nu(0)=\frac{1}{3Q^2}\left(\eta^{\mu\alpha}\eta^{\nu \beta}-\frac{q^\mu
 q^\alpha}{Q^2}\eta^{\nu\beta}-\eta^{\mu\alpha}
 \frac{q^\nu q^\beta}{Q^2}+\eta^{\mu\nu}
 \frac{q^\alpha
 q^\beta}{(Q^2)^2}\right) T_{\alpha\beta}\nonumber
 \\ -\frac{2}{15(Q^2)^2}\left(\eta^{\mu\nu}-\frac{q^\mu
 q^\nu}{Q^2}\right)q^\alpha q^\beta T_{\alpha\beta}+ \cdots\,,
 \eeq
from which one can read the coefficients of $T_{\mu\nu}$ in the OPE of
the current--current correlator. Although explicitly obtained here via a
lowest--order calculation in perturbation theory, these coefficients turn
out to be exactly as those (indirectly) computed at strong coupling, in
Sect. 3. To see that, let us specialize (\ref{theta}) to the high--energy
regime, where
 \beq
 q^\alpha q^\beta T_{\alpha\beta} \approx (q^-)^2T_{--}\approx
 2q^2 T_{--}\,. \eeq
and then take the thermal expectation value by using the expression
(\ref{theta}) for the average energy momentum--tensor
$\Theta_{\mu\nu}\equiv\langle T_{\mu\nu}\rangle$ in a {\em
strongly--coupled} SYM plasma. We thus obtain
 \beq i\int \rmd^4x \,\rme^{-iqx}
 \langle J^\mu(x)J^\nu(0)\rangle=
 \frac{\pi^2N^2T^4}{6Q^2}\left(n^{\mu}n^{\nu}-\frac{q\cdot n}{Q^2}q^\mu
 n^\nu - \frac{q\cdot n}{Q^2}q^\nu n^\mu + \frac{(q\cdot n)^2}{(Q^2)^2}
 \eta^{\mu\nu} \right) \nonumber
 \\ -\frac{q^2\pi^2N^2T^4}{15(Q^2)^2}\left(\eta^{\mu\nu}-\frac{q^\mu
 q^\nu}{Q^2}\right) \nonumber \\ =
 \frac{\pi^2N^2T^4}{6Q^2}\left(n_\mu -\frac{n\cdot q}{Q^2}
  q_\mu\right) \left(n_\nu -\frac{n\cdot q}{Q^2}q^\nu
 \right)
 +\frac{\pi^2N^2T^4q^2}{10(Q^2)^2}\left(\eta^{\mu\nu}-\frac{q^\mu
 q^\nu}{Q^2}\right)
 \,,
 \eeq
which is in full agreement with (\ref{R12LT}), as anticipated (recall
that $x=Q^2/2qT$). Normally, the OPE coefficients at strong coupling are
extracted by studying 3- and 4--point correlation functions. Our method
in Sect. 3 is more straightforward (though limited to the energy momentum
tensor) in that we do not have to compute higher point functions, but
only use the known value of $\langle T_{\mu\nu}\rangle$ at finite
temperature.

\section{High energy: the WKB approximation}
\setcounter{equation}{0}

In this Appendix, we shall construct approximate solutions to the
gravitational wave equations in the high energy regime at $k\gg K^3$. We
shall thus confirm and extend the results found in Sect. 4, which, we
recall, were valid only for $u\ll 1$.  The complete solutions will be
obtained by matching three different approximations, valid for different
values of $u$: the two limited solutions valid for $u\ll 1$ and near
$u=1$, respectively, and the WKB solution valid in the intermediate
region at $0<u\ll 1$. In this construction, the outgoing--wave condition
will be imposed near the black hole horizon at $u=1$, in conformity with
the original prescription in Refs. \cite{Son:2002sd,Son:2007vk}.

The general solutions valid for $u\ll 1$ have been already constructed in
Sect. 4. In the longitudinal sector, this is given by Eq.~(\ref{57}),
where the coefficient $c_2$ is fixed by the boundary condition at $u=0$,
with the result shown in Eq.~(\ref{c2}); as for $c_1$, this will be here
obtained by matching onto the solution near $u=1$, via the intermediate
WKB solution.

Consider now the solution near the horizon. For $k\gg K^3$ and $u\simeq
1$, Eq.~(\ref{ad}) simplifies to
  \beq\label{psiul} \psi^{\prime\prime}+\frac{k^2}{4(1-u)^2}\psi=0\,. \eeq
The solution which obeys the right, outgoing--wave, behaviour near $u=1$
reads
  \beq \psi(u)=c_3(1-u)^{\frac{1}{2}(1-ik)}\,.
   \label{60} \eeq
(The second independent solution $(1-u)^{\frac{1}{2}(1+ik)}$ must be
rejected since it would describe a wave coming out from the horizon,
i.e., a wave reflected by the black hole.)

Furthermore, in the intermediate region $u_0\ll u \ll 1$, with $u_0=
1/(4k^2)^{1/3}$, the `Schr\"odinger' wave equation reads
  \beq \psi^{\prime\prime}+
  \frac{k^2u}{(1-u^2)^2} \psi =0\,. \label{62}  \eeq
The WKB solution has the standard structure
$\psi(u)=\rme^{i\sigma_0(u)}/\sqrt{|\sigma_0^\prime|}$ with
 \beq\label{eikonal}
  \sigma_0(u)=\int_0^u \rmd u^\prime \sqrt{-V(u^\prime)} =\pm
 k\int_0^u\rmd u^\prime\frac{\sqrt{u^\prime}}{1-u^{\prime 2}} \nonumber \\
 =\pm \frac{k}{2}\left(\ln \frac{1+\sqrt{u}}{1-\sqrt{u}}- 2\arctan\sqrt{u}
 \right)\,.
 \eeq
Hence the general solution in this intermediate region reads
 \beq \psi(u)&\,=\,&\sqrt{\frac{1-u^2}{k\sqrt{u}}}\,
 \big[c_4F_k(u)\,+\,c_5F_{-k}(u)\big],\nn
 F_k(u)&\,\equiv\,& \left(\frac{1+\sqrt{u}}{1-\sqrt{u}}\right)^{ik/2}
 \,\rme^{-ik\arctan\sqrt{u}}\,.
 \label{66}
 \eeq

We can now determine the unknown coefficients by matching the previous
solutions in their common ranges of applicability. Comparing (\ref{60})
and (\ref{66}) near $u=1$ gives
 \beq c_5=0,\qquad c_3=c_4\sqrt{\frac{2}{k}}\
 \rme^{ik\ln 2-i\pi k/4}\,. \eeq
Then a comparison of (\ref{57}) and (\ref{66}) in the region $u\ll 1$ but
$u\gg u_0$  --- in this region $\psi(u)\simeq \sqrt{u}\,a(u)$ and $\xi\gg
1$, so one can use the asymptotic expansions for the Bessel functions in
Eq.~(\ref{57}) --- gives, after simple calculations,
  \beq\label{c34}
  c_1=-ic_2, \qquad c_4=\sqrt{\frac{3}{\pi}}\,c_2\,\rme^{-3i\pi/4}\,. \eeq
As anticipated, we have recovered the simple relation (\ref{c12}) between
$c_1$ and $c_2$ which implies that already the small--$u$ solution,
Eq.~(\ref{57}), is an outgoing wave, cf. Eq.~(\ref{71}).

Turning now to the transverse sector, where the small--$u$ solution was
given in Eq.~(\ref{Aismall}), we can similarly obtain the
(outgoing--wave) solution near $u=1$ as
  \beq\label{u1t} A_i(u)=c_3(1-u)^{-ik/2}\,, \eeq
and the corresponding WKB solution as (compare to Eq.~(\ref{66}))
 \beq
 A_i=\frac{c_4}{\sqrt{k\sqrt{u}}}\left(\frac{1+\sqrt{u}}{1-\sqrt{u}}\right)^{ik/2}
 \,\rme^{-ik\arctan\sqrt{u}}\,. \eeq
(We have anticipated that $c_5$ is set to zero after matching onto
Eq.~(\ref{u1t}).) The matching conditions then yield
 \beq c_1 = -ic_2, \qquad c_3=\frac{c_4}{\sqrt{q}}\,
 \rme^{ik\ln 2-i\pi k/4},\qquad
c_4=\sqrt{3/\pi}\, c_1\left(\frac{2k}{3}\right)^{{1}/{3}}
    \rme^{-5i\pi/12}\,. \eeq
One finally gets the same result at small $u$ as previously displayed in
Eq.~(\ref{81}).

\providecommand{\href}[2]{#2}
\begingroup\raggedright

\endgroup

\end{document}